\documentclass[aps,prx,twocolumn,showpacs]{revtex4-1}
\usepackage{bm}
\usepackage{mathrsfs}
\usepackage{amsmath}
\usepackage{amssymb}
\usepackage{graphicx}
\usepackage{amsfonts}
\usepackage{amsthm}
\usepackage{color}
\usepackage{dcolumn}
\usepackage{txfonts}
\usepackage[caption=false]{subfig}
\DeclareMathOperator{\erfc}{erfc}
\DeclareMathOperator{\Ai}{Ai}
\DeclareMathOperator{\Bi}{Bi}

\begin{document}

\title{Analytical Results for the Dynamics of Parabolic Level-Crossing Model}
\author{Chon-Fai Kam}
\author{Yang Chen}
\email{Corresponding author. Email: yangbrookchen@yahoo.co.uk}
\affiliation{Department of Mathematics, Faculty of Science and Technology,
University of Macau, Avenida da Universidade, Taipa, Macau, China}

\begin{abstract}
We study the dynamics of a two-level crossing model with a parabolic separation of the diabatic energies. The solutions are expressed in terms of the tri-confluent Heun equations --- the generalization of the confluent hypergeometric equations. We obtain analytical approximations for the state populations in terms of Airy and Bessel functions. Applicable expressions are derived for a large part of the parameter space. We also provide simple formulas which connect local solution in different time regimes. The validity of the analytical approximations is shown by comparing them to numerical simulations. 
\end{abstract}

\maketitle
\section{introduction}
Level crossing models are crucial for the understanding of non-adiabatic transitions in physics, chemistry and biology \cite{nakamura2012nonadiabatic}. The best known and probably most widely studied level-crossing model is the Landau-Zener model, which was formulated by Landau in 1932 for analyzing atomic collisions in both near-sudden \cite{landau1932theoryI} and near-adiabatic limits \cite{landau1932theory}, and was subsequently solved by Zener using parabolic cylinder functions \cite{zener1932non}. At around the same time, St{\"u}ckelberg derived a sophisticated tunneling formula based on analytical continuation of the semi-classical WKB solutions across the Stokes lines \cite{stuckelberg1932theory}, and Majorana derived the transition probability formula independently using integral representation of the survival amplitude in connection with the dynamics of a spin-1/2 in a time-varying magnetic field \cite{majorana1932atomi}. The Landau-Zener model assumes a constant coupling between bare states in the diabatic basis and a linearly varying separation of diabatic energies \cite{wittig2005landau}. The tunneling probability $\exp(-\zeta)$ in the Landau-Zener model only depends on a single dimensionless parameter $\zeta\equiv 2\pi f^2/(\hbar|(F_1-F_2)V|)$, where $f$ is the coupling matrix element in the diabatic basis, $F_1$ and $F_2$ are the slopes of the intersecting diabatic potential curves, and $V$ is the velocity of the perturbation variable, \textit{e.g.}, the relative collision velocity \cite{di2005majorana}. Since Landau, Zener, St{\"u}ckelberg and Majorana's pioneering works, the linear two-state model has been applied to atomic \cite{smirnov2003physics, nikitin2012theory} and molecular collisions \cite{child2010molecular}, atoms in intense laser fields \cite{delone1985atoms, kazantsev1990mechanical}, spin tunneling in molecular nano-magnets \cite{wernsdorfer1999quantum}, tunneling of Bose-Einstein condensates in accelerated optical lattices \cite{morsch2001bloch}, and optical tunneling in waveguide arrays \cite{khomeriki2005nonadiabatic}.

Although the Landau-Zener model has achieved great success over the last century, there are indeed cases where the assumption of linear crossing between the diabatic states breaks down. To be more precise, one may employ the St{\"u}ckelberg tunneling formula to the individual tunneling events \cite{stuckelberg1932theory, shevchenko2010landau}, as long as any pair of Landau-Zener crossings are well-separated from each other. In other words, for cases in which the crossing points merge together as a result of external electric or magnetic fields, the Landau-Zener linearization fails, and the linear-dependence of diabatic energies should be replaced by a parabolic one \cite{garraway1995wave}. The parabolic model was first introduced by Bikhovskii, Nikitin and Ovchinnikova in 1965 in the context of slow atomic collisions \cite{bikhovskii1965probability} and later re-evaluated by Delos and Thorson \cite{delos1972solution, delos1972studies} and Crothers \cite{crothers1975stueckelberg, crothers1976perturbed, crothers1977stueckelberg} in various limits. Two decades later, Shimshoni and Gefen incorporated environment-induced dissipation and dephasing into the parabolic model \cite{shimshoni1991onset}. Suominen derived analytical approximations for the final state populations \cite{suominen1992parabolic}, and applied the results to the dynamics of cold atoms in magnetic traps \cite{pakarinen2000atomic}. In the same period, Zhu and Nakamura derived an exact formula for the scattering matrices in terms of a convergent infinite series, in which the coefficients satisfy a five-term recursion relation \cite{zhu1992two, zhu1992twoII, zhu1993two, zhu1994two, zhu2001nonadiabatic}. Nakamura and co-workers applied the results to laser assisted surface ion neutralization \cite{teranishi1997semiclassical} and the laser-controlled photochromism in functional molecules \cite{tamura2006laser}. Over the last decade, Lehto incorporated super-parabolic level-glancing effects into the parabolic model \cite{lehto2012superparabolic, lehto2013zhu}, and studied complete population inversion due to phase-jump couplings \cite{lehto2016two}. Zhang and co-workers described the population dynamics of driven dipolar molecules in the parabolic level-glancing model in terms of the confluent Heun functions \cite{zhang2016analytic}. Recently, the parabolic model has found applications in topological systems, including the interband tunneling of fermionic atoms near the merging transition of Dirac cones in tunable honeycomb optical lattices \cite{fuchs2012interband}, and the interband tunneling of two-dimensional electrons near the topological transition in type II Weyl semimetals \cite{malla2019high}.

The dynamics of the parabolic level-crossing model, probably unknown to most physicists, may be written in terms of the tri-confluent Heun function \cite{ronveaux1995heun}, which is derived from the general Heun function \cite{heun1888theorie} by the coalescence of three finite regular singularities with infinity. Unlike the parabolic cylinder function appearing in the Landau-Zener model, comprehensive information on the asymptotic behavior of confluent Heun functions, in general, does not exist. The main difficulty is due to the fact that the Heun functions do not possess any integral representations in terms of simpler special functions. Hence, Majorana's integral representation method is not applicable for determining the transition probability at infinity. Nevertheless, in the subsequent sections, we will show that valuable analytical approximations which only involves ordinary special functions can still be obtained in a large part of parameter space.

The paper is organized as follows: In Section \ref{II}, we introduce the parabolic level-crossing model in the context of a two-level atom driven by a classical laser field, in which the laser detuning varies quadratically with time, and the Rabi frequency at resonance is time-independent. We express the final state population in terms of a single Stokes multiplier. In Section \ref{III}, we derive analytical approximations for the transition amplitudes in both short- and long-time regimes. The validity of the analytical approximations is tested via comparisons with numerical results. We also discuss a way to connect analytical solutions in different time regimes. Finally, in Section \ref{V}, we conclude our studies, and discuss extensions of the results to other level-crossing models.

\section{The Parabolic Model}\label{II}
Within the rotating-wave approximation, the wave amplitudes of a two-level atom dipole-interacting with a classical electric field are governed by the following coupled equations (details are provided in Appendix \ref{A})
\begin{equation}
i\frac{da_1}{dt}=-\frac{\Delta}{2}a_1+fa_2,\:
i\frac{da_2}{dt}=f^*a_1+\frac{\Delta}{2}a_2,
\end{equation}
where $\Delta$ is the laser detuning, and $f$ is the Rabi frequency at resonance. After applying the transformations $a_1=C_1e^{\frac{i}{2}\int_0^t\Delta ds}$ and $a_2=C_2e^{-\frac{i}{2}\int_0^t\Delta ds}$, the coupled equations becomes
\begin{equation}
    i\frac{dC_1}{dt}=fe^{-i\int_0^t\Delta ds}C_2,\:
    i\frac{dC_2}{dt}=f^*e^{i\int_0^t\Delta ds}C_1,
\end{equation}
or equivalently
\begin{subequations}
\begin{gather}
    \frac{d^2C_1}{dt^2}+\left(i\Delta-\frac{\dot{f}}{f}\right)\frac{dC_1}{dt}+|f|^2C_1=0,\label{TwoLevelProblem2}\\
    \frac{d^2C_2}{dt^2}-\left(i\Delta+\frac{\dot{f}^*}{f^*}\right)\frac{dC_2}{dt}+|f|^2C_2=0,\label{TwoLevelProblem1}
\end{gather}
\end{subequations}
where $C_1$ and $C_2$ satisfy the normalization condition $|C_1|^2+|C_2|^2=1$. The problem of non-adiabatic  transition is to determine the transition probability $|C_1(\infty)|^2$, subjected to the conditions $|C_1(-\infty)|=0$ and  $|C_2(-\infty)|=1$. Using the change of variable $C_1=U_1\exp\{-\frac{1}{2}\int_0^tp ds\}$, where $p\equiv i\Delta-\dot{f}/f$, we obtain the Schr\"{o}dinger form of Eq.\:\eqref{TwoLevelProblem2}
\begin{subequations}
\begin{gather}
    \frac{d^2U_1}{dt^2}+J(t)U_1=0,\\
    J(t)=|f|^2-\frac{1}{2}\frac{d}{dt}\left(i\Delta-\frac{\dot{f}}{f}\right)-\frac{1}{4}\left(i\Delta-\frac{\dot{f}}{f}\right)^2.
\end{gather}
\end{subequations}
In the conventional Landau-Zener model, the laser detuning varies linearly with time, $\nu\equiv\alpha t$, and the the Rabi frequency at resonance is time-independent, which yields
\begin{equation}\label{Weber}
    \frac{d^2U_1}{dt^2}+\left(|f|^2-\frac{i\alpha}{2}+\frac{\alpha^2t^2}{4}\right)U_1=0.
\end{equation}
Eq.\:\eqref{Weber} becomes the parabolic cylinder equation $U_1''+(n+\frac{1}{2}-\frac{1}{4}z^2)U_1=0$ via the change of variables $z\equiv e^{-i\pi/4}\alpha^{1/2}t$ and $n\equiv i|f|^2/\alpha$. The transition probability $|C_1(\infty)|^2$ can be obtained from the asymptotic expansions of the parabolic cylinder function (details are provided in Appendix \ref{B}).

In contrast to the conventional Landau-Zener model, the laser detuning in the parabolic model varies quadratically with time, $\Delta\equiv\alpha t+\frac{1}{2}\beta t^2$, and the Rabi frequency at resonance is time-independent, which yields
\begin{equation}\label{TriConflueneHeun}
    \frac{d^2U_1}{dt^2}+\left(|f|^2-\frac{i\alpha}{2}-\frac{i\beta t}{2}+\frac{\alpha^2t^2}{4}+\frac{\alpha\beta t^3}{4}+\frac{\beta^2t^4}{16}\right)U_1=0.
\end{equation}
Eq.\:\eqref{TriConflueneHeun} may be transformed into the canonical form of tri-confluent Heun equation. Let us perform the transformation $z\equiv h^{-1}(t+\alpha/\beta)$ with $h^6=-\frac{9}{4}\lambda^{-2}$ and $\lambda\equiv \beta/4$, then Eq.\:\eqref{TriConflueneHeun} becomes the second canonical form of the tri-confluent Heun equation (\textit{THE$_2$} equation) \cite{ronveaux1995heun}
\begin{subequations}
\begin{align}
    &\frac{d^2U_1}{dz^2}+\left(\mu-\frac{\xi^2}{4}+\nu z-\frac{3}{2}\xi z^2-\frac{9}{4}z^4\right)U_1=0,\\
    \mu&\equiv\left(|f|^2+\frac{\alpha^4}{16\beta^2}\right)h^2+\frac{\xi^2}{4},\nu=3,\xi\equiv-3\left(\frac{\alpha}{h\beta}\right)^2,
\end{align}
\end{subequations}
which may be transformed into the first canonical form of the tri-confluent Heun equation (\textit{THE$_1$} equation) \cite{ronveaux1995heun}
\begin{equation}
\frac{d^2V_1}{dz^2}+(\xi+3z^2)\frac{dV_1}{dz}+(\mu+(\nu-3)z)V_1=0,
\end{equation}
via the transformation $U_1=e^{-\frac{1}{2}(z^3+\xi z)}V_1$. To be precise and for later convenience, we define $h\equiv e^{i\pi/6}(3/2\lambda)^{1/3}$ for $\lambda>0$, and define $\mathcal{L}U_1(\mu,\nu,\xi;z)=U_1(\omega^4\mu,\omega^3\nu,\omega^2\xi;\omega z)$ for $\omega=e^{i\pi/3}$. 

Similar to Ziner's approach to the linear level-crossing model \cite{zener1932non}, the transition probability may be derived from the asymptotic expansions of the tri-confluent Heun function at different sectors in the complex plane. The \textit{THE}$_2$ equation has two independent solutions $T_1(\mu,\nu,\xi;z)$ and $T_2(\mu,\nu,\xi;z)$, where $T_1(\mu,\nu,\xi;z)$ has the following asymptotic expansion in the sector $|\arg{z}|<\frac{\pi}{2}$ \cite{ronveaux1995heun}
\begin{equation}
    T_1(\mu,\nu,\xi;z)=e^{-\frac{1}{2}(z^3+\xi z)}z^{\frac{\nu}{3}-1}\sum_{k\geq 0}a_k(\mu,\nu,\xi)z^{-k},
\end{equation}
and $T_2(\mu,\nu,\xi;z)$ has the following asymptotic expansion in the sector $\frac{\pi}{2}<\arg z<\frac{3\pi}{2}$ \cite{ronveaux1995heun}
\begin{align}\label{T2}
    T_2(\mu,\nu,\xi;z)&=T_1(\mu,-\nu,\xi;-z)\nonumber\\
    &=e^{\frac{1}{2}(z^3+\xi z)}z^{-\frac{\nu}{3}-1}\sum_{k\geq 0}(-1)^ka_k(\mu,-\nu,\xi)z^{-k},
\end{align}
where $a_0(\mu,\nu,\xi)=1$, $a_1(\mu,\nu,\xi)=-\mu/3$, $a_2(\mu,\nu,\xi)=\frac{1}{18}(\mu^2+\xi(\nu-3))$ and $a_k(\mu,\nu,\xi)$ satisfies the following four-term recursion relation \cite{ronveaux1995heun}
\begin{gather}
3(k+3)a_{k+3}+\mu a_{k+2}+\xi(k+2-\nu/3)a_{k+1}\nonumber\\
+(k+1-\nu/3)(k+2-\nu/3)a_k=0.
\end{gather}
For $t\rightarrow -\infty$, we have $z=-h^{-1}|t+\alpha/\beta|$ and $\arg z=\frac{5\pi}{6}$. Hence, we may use the solution $T_2(\mu,\nu,\xi;z)=T_1(\mu,-\nu,\xi;-z)$, so that
\begin{align}
    U_1(t\rightarrow -\infty)&=A_1T_1(\mu,-\nu,\xi;h^{-1}|t+\alpha/\beta|)\nonumber\\
    &\approx A_1(\beta/6)^{-\frac{2}{3}}e^{-i(\frac{\beta}{12}t^3+\frac{\alpha}{4}t^2-\frac{1}{6}\frac{\alpha^3}{\beta^2}-\frac{\pi}{3})}t^{-2}.
\end{align}
Using the relation $C_1=U_1\exp\{-\frac{i}{2}\int_0^t(\alpha s+\frac{\beta}{2} s^2)ds\}$, we obtain
\begin{subequations}
\begin{align}\label{TriHeunExact1}
    C_1(t\rightarrow -\infty)&\approx A_1(\beta/6)^{-\frac{2}{3}}e^{-i(\frac{\beta}{6}t^3+\frac{\alpha}{2}t^2-\frac{1}{6}\frac{\alpha^3}{\beta^2}-\frac{\pi}{3})}t^{-2},\\
    \dot{C}_1(t\rightarrow -\infty)&\approx -3i(\beta/6)^{\frac{1}{3}}A_1e^{-i(\frac{\beta}{6}t^3+\frac{\alpha}{2}t^2-\frac{1}{6}\frac{\alpha^3}{\beta^2}-\frac{\pi}{3})}.\label{TriHeunExact2}
\end{align}
\end{subequations}
The constant $A_1$ is determined by $|\dot{C}_1|=|f\dot{C}_2|=|f|$, which yields $A_1=\frac{|f|}{3}(\frac{\beta}{6})^{-\frac{1}{3}}$. The large $|t|$ solutions Eqs.\:\eqref{TriHeunExact1} and \eqref{TriHeunExact2} can also be obtained from the method of direct integration (details are provided in Appendix \ref{C}). For $t\rightarrow\infty$, we have $z=h^{-1}(t+\alpha/\beta)$ and $\arg z=-\pi/6$. Hence, Eq.\:\eqref{T2},  the asymptotic expansion for $T_2(\mu,\nu,\xi;z)$ may not be used. In order to evaluate $T_2(\mu,\nu,\xi;z)$ for $t\rightarrow \infty$, we have to use the connection formula \cite{ronveaux1995heun}
\begin{equation}
    T_2(\mu,\nu,\xi;z)=\mathcal{L}T_1(\mu,\nu,\xi;z)-\mathcal{L}C(\mu,\nu,\xi)T_1(\mu,\nu,\xi;z),
\end{equation}
where $\mathcal{L}C(\mu,\nu,\xi)\equiv C(\omega^4\mu,-\nu,\omega^2\xi)$. $C(\mu,\nu,\xi)$ is the Stokes multiplier which connects the asymptotic expansions of $U_1(z)$ at different sectors, and is an entire function of $\mu$, $\nu$ and $\xi$ \cite{sibuya1975global}. For $t\rightarrow\infty$, we have $\arg(h^{-1}(t+\alpha/\beta))=-\frac{\pi}{6}$ and $\arg(\omega h^{-1}(t+\alpha/\beta))=\frac{\pi}{6}$. Hence, we may use the asymptotic expansion of $T_1(\mu,\nu,\xi;z)$ and obtain
\begin{align}
    U_1(t\rightarrow\infty)&\approx A_1[(\beta/6)^{-\frac{2}{3}}e^{-i(\frac{\beta}{12}t^3+\frac{\alpha}{4}t^2-\frac{1}{6}\frac{\alpha^3}{\beta^2}+\frac{\pi}{3})}t^{-2}\nonumber\\
    &-C(\omega^4\mu,-\nu,\omega^2\xi)e^{i(\frac{\beta}{12}t^3+\frac{\alpha}{4}t^2-\frac{1}{6}\frac{\alpha^3}{\beta^2})}].
\end{align}
Using the relation $C_1=U_1e^{-i(\frac{\beta}{12}t^3+\frac{\alpha}{4}t^2)}$, we obtain
\begin{equation}
    C_1(t\rightarrow\infty)\approx -A_1C(\omega^4\mu,-\nu,\omega^2\xi)e^{-i\frac{\alpha^3}{6\beta^2}},
\end{equation}
which yields $|C_1(\infty)|^2=|A_1C(\omega^4\mu,-\nu,\omega^2\xi)|^2$. Hence, the final transition probability $|C_1(\infty)|^2$ depends only on the Stokes multiplier $C(\mu,\nu,\xi)$. However, it is in general not an easy task to obtain exact formulas for the Stokes multipliers. Although Zhu and Nakamura derived an exact formula for the Stokes multipliers in terms of a sophisticated infinite series generated by a five-term recursion relation \cite{zhu1992stokes}, a compact formula for the final transition probability which is similar to the Landau-Zener formula has not yet existed. In the following section, we derive concise and explicit expressions for the transition dynamics in the parabolic model, and provide an analytical approximation for connecting solutions of the transition amplitude in different time regimes.

\section{Analytical Approximations for the transition amplitude}\label{III}
In the last section, we have shown that the dynamics of a two-level atom dipole-interacting with an off-resonant classical electric field with constant amplitude and parabolic detuning can be solved in terms of the tri-confluent Heun functions. We discussed the relationship between the final transition probability and the Stoke multipliers which connect asymptotic expansions of the tri-confluent Heun functions. However, due to mathematical difficulties involved, it is better to develop analytical approximations to the transition amplitudes, rather than to solve the connection problem rigorously.

To begin with, let us rewrite Eq.\:\eqref{TriConflueneHeun}, the differential equation which governs $U_1$, as
\begin{equation}\label{Tau}
    \frac{d^2U_1}{d\tau^2}+\left[|f|^2-\frac{i\beta\tau}{2}+\frac{\beta^2}{16}\left(\tau^2-\frac{\alpha^2}{\beta^2}\right)^2\right]U_1=0,
\end{equation}
where $\tau\equiv t+\alpha/\beta$. From Eq.\:\eqref{Tau}, we see that the sign of $\alpha/\beta$ does not alter the nature of the equation. Hence, without loss of generality, we may assume that $\alpha/\beta$ is a positive number. Eq.\:\eqref{Tau} can also be rewritten as
\begin{equation}\label{GeneralTau}
    \frac{d^2U_1}{d\tau^2}+\left(|f|^2+\frac{\alpha^4}{16\beta^2}-\frac{i\beta\tau}{2}-\frac{\alpha^2\tau^2}{8}+\frac{\beta^2\tau^4}{16}\right)U_1=0.
\end{equation}

\begin{figure}[tbp]
\begin{center}
\includegraphics[width=0.75\columnwidth]{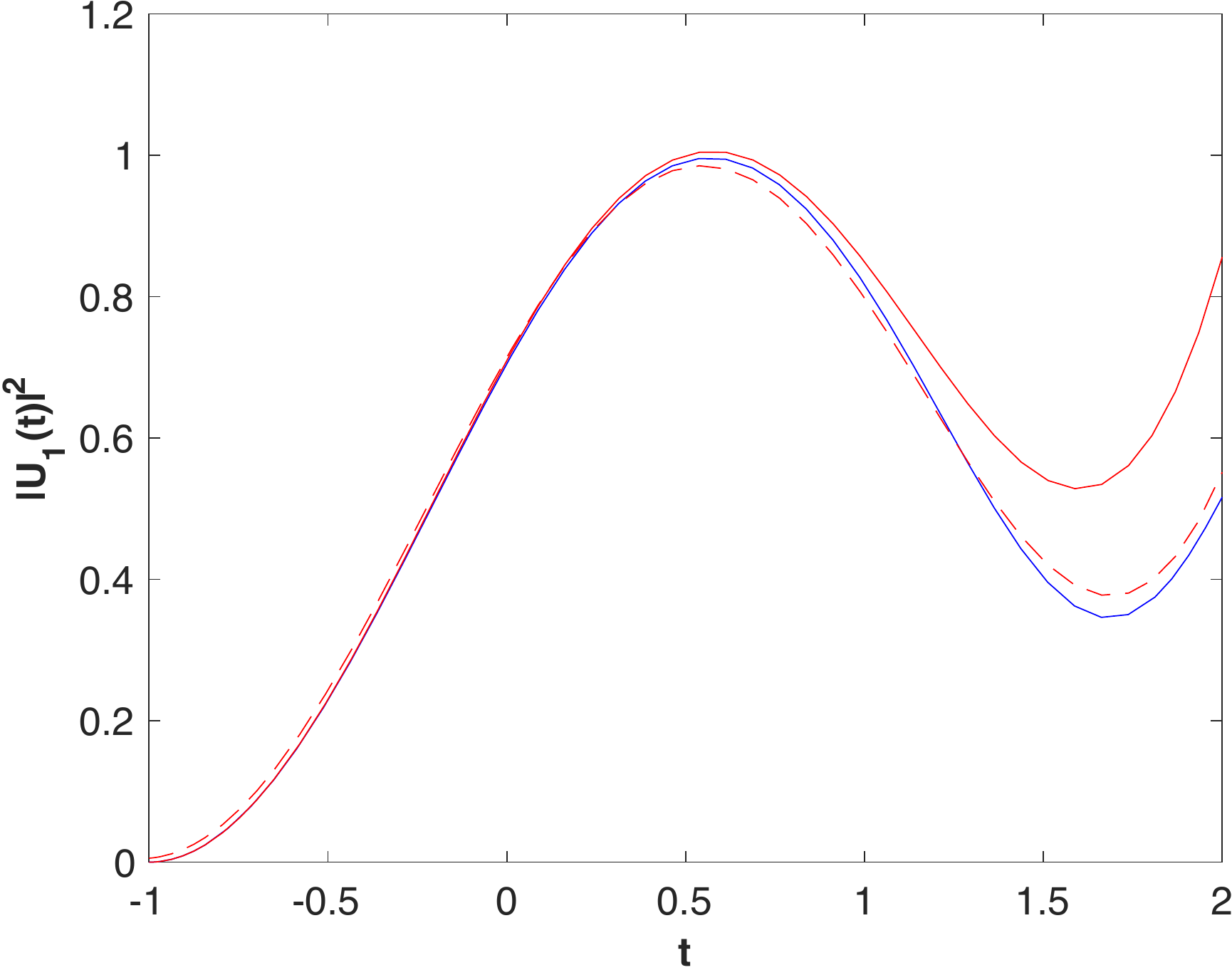}
\end{center}
\caption{The transition probability $|U_1(t)|^2$ subjected to the initial conditions $|U_1(t_i)|=0$ and $|\dot{U}_1(t_i)|=|f|$. Exact solution of Eq.\:\eqref{TwoLevelProblem2} is depicted in blue, approximate solution of the homogeneous equation $\bar{U}^{\prime\prime}_1=z\bar{U}_1$ is depicted in red solid line, and the analytical approximation to Eq.\:\eqref{HeunFromAiry}, which is given by Eq.\:\eqref{HeunAiryAnalyticalApproximation}, is depicted in red dashed line. Here, $t_i=-1$, $t_f=2$, $|f|=1$, $\alpha=0.2$ and $\beta=2$.}
\label{fig:HeunAiry}
\end{figure}For large $\alpha/\beta$, we compare the last two terms in Eq.\:\eqref{GeneralTau}. The short time regime is defined by $\beta^2\tau^4/16\lessapprox\alpha^2\tau^2/8$, \textit{i.e.}, $|t|\lessapprox t^*\approx \alpha/\beta$; whereas the long time regime is defined by $\beta^2\tau^4/16\gtrapprox\alpha^2\tau^2/8$, \textit{i.e.}, $|t|\gtrapprox t^*$. In the short time regime, the quadratic term is dominant in Eq.\:\eqref{GeneralTau}. Hence, the transition dynamics can be described by the Landau-Zener formula, and solved via the parabolic cylinder functions. But for small $\alpha/\beta$, we should compare the third and the last terms in Eq.\:\eqref{GeneralTau}. Hence, the short time regime is defined by $\beta^2\tau^4/16\lessapprox|\beta\tau|/2$, \textit{i.e.}, $|\tau|\lessapprox 2|\beta|^{-1/3}$; whereas the long time regime is defined by $|\tau|\gtrapprox2|\beta|^{-1/3}$. In the short time regime, the linear term is dominant in Eq.\:\eqref{GeneralTau}. Hence, the transition dynamics is solved via the Airy functions, which is different from that obtained from the Landau-Zener formula. In the long time regime where the quartic term is dominant in Eq.\:\eqref{GeneralTau}, the transition dynamics can be solved via the Bessel functions. 

In the following subsections, we discuss in detail the analytical approximations for the transition amplitude in different time regimes.

\subsection{Dynamics in the short-time regime for small $\alpha/\beta$}
In this subsection, we analyze the dynamics for the transition amplitude in the time regime $|t|\lessapprox t^*$, so that $\beta^2\tau^4/16\lessapprox\alpha^2\tau^2/8$. We require $|\alpha|\lessapprox\sqrt{2}\beta^{2/3}$, so that $\alpha^2\tau^2/8\lessapprox|\beta\tau|/2$ is satisfied. We may rewrite Eq.\:\eqref{GeneralTau} as
\begin{equation}\label{LargeTPerturbative}
    \frac{d^2U_1}{d\tau^2}+\left(|f|^2+\frac{\alpha^4}{16\beta^2}-\frac{i\beta\tau}{2}\right)U_1=\left(\frac{\alpha^2\tau^2}{8}-\frac{\beta^2\tau^4}{16}\right)U_1,
\end{equation}
where the right-hand side of Eq.\:\eqref{LargeTPerturbative}
is regarded as a known function of $\tau$. Let us denote $z=(\frac{i\beta}{2})^{\frac{1}{3}}[\tau+\frac{2i}{\beta}(|f|^2+\frac{\alpha^4}{16\beta^2})]$, then we obtain $\tau=(\frac{i\beta}{2})^{-\frac{1}{3}}z-\frac{2i}{\beta}(|f|^2+\frac{\alpha^4}{16\beta^2})$ and
\begin{align}\label{HeunFromAiry}
    \frac{d^2U_1}{dz^2}-zU_1&=\left\{\frac{1}{4}\left(\frac{i\beta}{2}\right)^{2/3}[(z-z_+)(z-z_-)]^2+\frac{\alpha^4}{16\beta^2}\right\}U_1\nonumber\\
    &\equiv \pi_4(z)U_1,
\end{align}
where $z_\pm\equiv(\frac{i\beta}{2})^{1/3}(2i\gamma\pm 1)\frac{\alpha}{\beta}$ and $\gamma\equiv \frac{1}{\alpha}(|f|^2+\frac{\alpha^4}{16\beta^2})$. From Eq.\:\eqref{HeunFromAiry}, we see that the homogeneous equation $\bar{U}^{\prime\prime}_1=z\bar{U}_1$ is solved by the Airy functions $\Ai(z)$ and $\Bi(z)$. To be specific, we define the cubic root of $\frac{i\beta}{2}$ as $e^{\pm i\frac{\pi}{6}}(\frac{|\beta|}{2})^{\frac{1}{3}}$ for $\beta=\pm|\beta|$, so that $\arg z\in(\frac{\pi}{6},\frac{7\pi}{6})$ for $\beta>0$ and $\arg z\in(-\frac{7\pi}{6},-\frac{\pi}{6})$ for $\beta<0$. The particular solution of Eq.\:\eqref{HeunFromAiry} is given by
\begin{align}
    U_{1p}(z)&=-\Ai(z)\int_{z_0}^z\frac{\Bi(\zeta)\pi_4(\zeta)\bar{U}_1(\zeta)}{\mathscr{W}\{\Ai(\zeta),\Bi(\zeta)\}}d\zeta\nonumber\\
    &+\Bi(z)\int_{z_0}^z\frac{\Ai(\zeta)\pi_4(\zeta)\bar{U}_1(\zeta)}{\mathscr{W}\{\Ai(\zeta),\Bi(\zeta)\}}d\zeta,
\end{align}
where $\mathscr{W}\{\Ai(\zeta),\Bi(\zeta)\}=1/\pi$ is the Wronskian of $\Ai(\zeta)$ and $\Bi(\zeta)$. As the homogeneous solution $\bar{U}_1(z)$ is a linearly combination of $\Ai(z)$ and $\Bi(z)$, we need to evaluate integrals of the form $\int z^ny^2dz$ and $\int z^n y_1y_2dz$, where $y$ refers to $\Ai(z)$ or $\Bi(z)$ respectively. After integration, Eq.\:\eqref{HeunFromAiry} is solved by (details can be found in Appendix \ref{AiryAppendix})
\begin{equation}\label{HeunAiryAnalyticalApproximation}
    U_1(z)\approx \left(1-\frac{1}{2}\frac{dG(z)}{dz}\right)\bar{U}_1(z)+G(z)\bar{U}^\prime_1(z),
\end{equation}
where $\bar{U}_1(z)\equiv c_1\Ai(z)+c_1\Bi(z)$ with $c_1$ and $c_2$ being arbitrary constants, and
\begin{align}\label{HeunAiryG}
    G(z)&=\frac{1}{4}\left(\frac{i\beta}{2}\right)^{2/3}\left[\frac{z^4}{9}-\frac{2az^3}{7}+\frac{a^2+2b}{5}z^2\right.\nonumber\\
    &\left.+\left(\frac{4}{9}-\frac{2ab}{3}\right)z+b^2-\frac{6a}{7}\right]+\frac{\alpha^4}{16\beta^2},
\end{align}
where $a\equiv z_++z_-$ and $b\equiv z_+z_-$. In Fig.\:\ref{fig:HeunAiry}, we compare the analytical approximation of the transition probability with the numerical solution. The result shows that the transition probability is well-described by Eqs.\:\eqref{HeunAiryAnalyticalApproximation} and \eqref{HeunAiryG} in the short-time regime for small $\alpha/\beta$.

\subsection{Dynamics in the short-time regime for large $\alpha/\beta$}
In this subsection, we analyze the dynamics for the transition amplitude in the time regime $|t|\lessapprox t^*$, so that $\beta^2\tau^4/16\lessapprox\alpha^2\tau^2/8$. We require $|\alpha|\gtrapprox\sqrt{2}\beta^{2/3}$, so that $\alpha^2\tau^2/8\gtrapprox|\beta\tau|/2$ is satisfied. In this regime, it is better to use the original equation which governs $U_1$. From Eq.\:\eqref{TriConflueneHeun}, we obtain
\begin{equation}\label{WeberSimple}
    \frac{d^2U_1}{dt^2}+\left(|f|^2-\frac{i\alpha}{2}+\frac{\alpha^2t^2}{4}\right)U_1\approx\left(\frac{i\beta t}{2}-\frac{\alpha\beta t^3}{4}\right)U_1,
\end{equation}
\begin{figure}[tbp]
	\subfloat[$t_i=-1$, $t_f=2$\label{sfig:ParabolicCylinderWithCorrection}]{%
	\includegraphics[width=0.75\columnwidth]{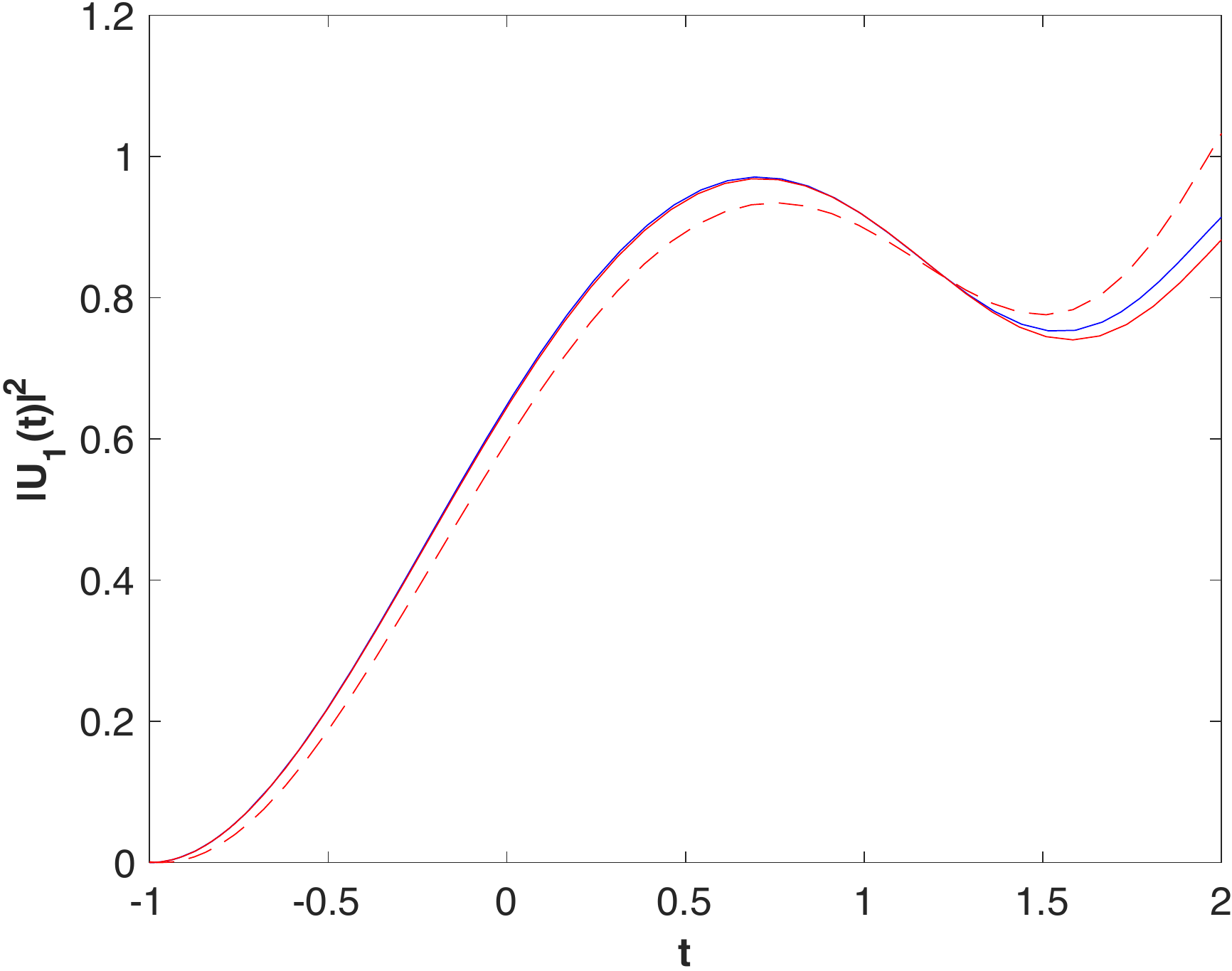}%
	}\hfill
	\subfloat[$t_i=-10$, $t_f=10$\label{sfig:ParabolicCylinderWithoutCorrection}]{%
 	 \includegraphics[width=0.75\columnwidth]{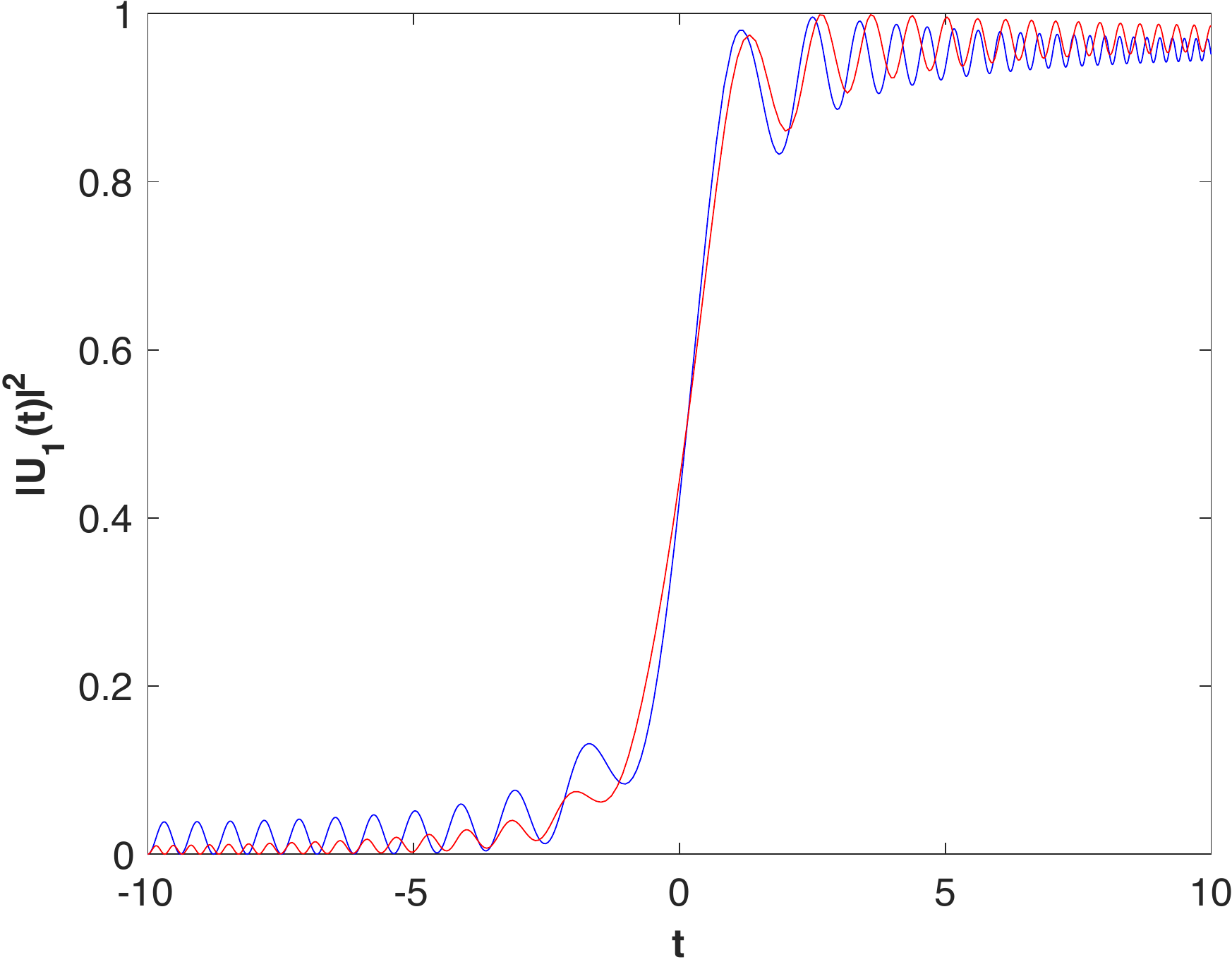}%
	}
\caption{The transition probability $|U_1(t)|^2$ subjected to the initial conditions $|U_1(t_i)|=0$ and $|\dot{U}_1(t_i)|=|f|$. Exact solution of Eq.\:\eqref{TwoLevelProblem2} is depicted in blue, approximate solution of the homogeneous equation $\bar{U}^{\prime\prime}_1=(a+\frac{1}{4}z^2)\bar{U}_1$ is depicted in red solid line, and the analytical approximation to Eq.\:\eqref{WeberAndHeun}, which is given by Eq.\:\eqref{HeunParabolicCylinder}, is depicted in red dashed line. Here, $|f|=1$, $\alpha=2$, and $\beta=0.2$.}
\label{fig:HeunParabolicCylinder}
\end{figure}where the right-hand side of Eq.\:\eqref{WeberSimple} is regarded as a known function of $t$. Let us denote $z=e^{-i\pi/4}\alpha^{1/2}t$, then we obtain
\begin{equation}\label{WeberAndHeun}
     \frac{d^2U_1}{dz^2}-\left(a+\frac{z^2}{4}\right)U_1\approx \left(-\frac{z}{2z^*}+\frac{z^3}{4z^*}\right)U_1\equiv\pi_3(z)U_1,
\end{equation}
where $a\equiv-i|f|^2/\alpha-1/2$ and $z^*\equiv e^{-i\pi/4}\alpha^{3/2}/\beta$. The homogeneous equation $\bar{U}^{\prime\prime}_1=(a+\frac{1}{4}z^2)\bar{U}_1$ is solved by the parabolic cylinder functions $U(a,\pm z)$ and $U(-a,\pm iz)$. We choose the pair of linearly independent solutions as $\{U(a,z),U(-a,\mp iz\}$, where the plus and minus signs are for $\alpha>0$ and $\alpha<0$ respectively. The particular solution of Eq.\:\eqref{WeberAndHeun} is given by
\begin{align}
    U_{1p}(z)&= -U(a,z)\int_{z_0}^z\frac{U(-a,\mp i\zeta)\pi_3(\zeta)\bar{U}_1(\zeta)}{\mathscr{W}\{U(a,\zeta),U(-a,\mp i\zeta)\}}d\zeta\nonumber\\
    &+U(-a,\mp iz)\int_{z_0}^z\frac{U(a,\zeta)\pi_3(\zeta)\bar{U}_1(\zeta)}{\mathscr{W}\{U(a,\zeta),U(-a,\mp i\zeta)\}}d\zeta,
\end{align}
where $\mathscr{W}\{U(a,\zeta),U(-a,\mp i\zeta)\}=\pm ie^{\mp i\pi(\frac{a}{2}+\frac{1}{4})}$ is the Wronskian of $U(a,\zeta)$ and $U(-a,\mp i\zeta)$. As the homogeneous solution $\bar{U}_1(z)$ is a linearly combination of $U(a,z)$ and $U(-a,\mp iz)$, we need to evaluate integrals of the form $\int z^ny^2dz$ and $\int z^ny_1y_2dz$, where $y$ refers to $U(a,\zeta)$ or $U(-a,\mp i\zeta)$ respectively. After integration, Eq.\:\eqref{WeberAndHeun} can still be solved by (details can be found in Appendix \ref{AiryAppendix})
\begin{align}\label{HeunParabolicCylinder}
    U_1(z)&\approx \left(1-\frac{1}{2}\frac{dG(z)}{dz}\right)\bar{U}_1(z)+G(z)\bar{U}_1^\prime(z),
\end{align}
where $\bar{U}_1(z)\equiv c_1 U(a,z)+c_2 U(-a,\mp iz)$ with $c_1$ and $c_2$ being arbitrary constants, and $G(z)=-(6+8a-z^2)/(6z^*)$. As we can see from Fig.\:\ref{sfig:ParabolicCylinderWithCorrection}, the Landau-Zener solution $|\bar{U}_1(z)|^2=|c_1 U(a,z)+c_2 U(-a,\mp iz)|^2$ coincides with the exact result of the transition probability obtained from Eq.\:\eqref{Tau}, but the analytical approximation Eq.\;\eqref{HeunParabolicCylinder} does not provide better result than the Landau-Zener solution. In Fig.\:\ref{sfig:ParabolicCylinderWithoutCorrection}, we compare the transition probability obtained from $|\bar{U}_1(z)|^2$ with the exact solution, and the result shows that the final transition probability is well-described by the Landau-Zener formula in the whole-time range for large $\alpha/\beta$.

\subsection{Dynamics in the long-time regime}
In this subsection, we analyze the dynamics for the transition amplitude in the time regime $|t|\gtrapprox t^*$, so that $\beta^2\tau^4/16\gtrapprox\alpha^2\tau^2/8$. We may rewrite Eq.\:\eqref{GeneralTau} as
\begin{align}\label{BesselOneSix}
    \frac{d^2U_1}{d\tau^2}+\frac{\beta^2\tau^4}{16}U_1
    &=\left(-|f|^2-\frac{\alpha^4}{16\beta^2}+\frac{i\beta\tau}{2}+\frac{\alpha^2\tau^2}{8}\right)U_1\nonumber\\
    &\equiv\pi_2(\tau)U_1,
\end{align}
where the right-hand side of Eq.\:\eqref{BesselOneSix} is regarded as a known function of $\tau$. The homogeneous equation $\bar{U}_1^{\prime\prime}=-\frac{1}{16}\beta^2\tau^4\bar{U}_1$ is solved by the Bessel functions of order $1/6$, that is, $w_1\equiv\sqrt{\tau}J_{1/6}(\beta \tau^3/12)$ and $w_2\equiv\sqrt{\tau}J_{-1/6}(\beta \tau^3/12)$. The particular solution of Eq.\:\eqref{BesselOneSix} is given by
\begin{align}\label{ParticularBesselOneSix}
    U_{1p}(\tau)&= -w_1(\tau)\int_{\tau_0}^\tau\frac{w_2(\tau^\prime)\pi_2(\tau^\prime)\bar{U}_1(\tau^\prime)}{\mathscr{W}\{w_1(\tau^\prime),w_2(\tau^\prime)\}}d\tau^\prime\nonumber\\
    &+w_2(\tau)\int_{\tau_0}^\tau\frac{w_1(\tau^\prime)\pi_2(\tau^\prime)\bar{U}_1(\tau^\prime)}{\mathscr{W}\{w_1(\tau^\prime),w_2(\tau^\prime)\}}d\tau^\prime,
\end{align}
where $\mathscr{W}\{w_1(\tau),w_2(\tau)\}=-3/\pi$ is the Wronskian of $w_1(\tau)$ and $w_2(\tau)$. As the homogeneous solution $\bar{U}_1$ is a linearly combination of $w_1$ and $w_2$, we need to evaluate integrals of the form $I_n\equiv \int \tau^n w^2 d\tau$ and $J_n\equiv \int \tau^n w_1w_2 d\tau$, where $w$ refers to $w_1$ or $w_2$ respectively. After integration, Eq\:\eqref{ParticularBesselOneSix} is solved by (details can be found in Appendix \ref{AiryAppendix})
\begin{align}\label{HeunBesselAnalytic}
    U_1(\tau)&\approx \left(1-\frac{1}{2}\frac{dG(\tau)}{d\tau}\right)\bar{U}_1(\tau)+G(\tau)\bar{U}_1^\prime(\tau),
\end{align}
where $\bar{U}_1(\tau)=c_1 \sqrt{\tau}J_{1/6}\left(\beta \tau^3/12\right)+c_2 \sqrt{\tau}J_{-1/6}\left(\beta \tau^3/12\right)$ with $c_1$ and $c_2$ being arbitrary constants, and
\begin{align}
    G(\tau)&=\left(|f|^2+\frac{\alpha^4}{16\beta^2}\right)\frac{\tau^3}{3}{}_2F_3\left(1,\frac{5}{6};\frac{7}{6},\frac{8}{6},\frac{9}{6};-\left(\frac{\beta\tau^3}{12}\right)^2\right)\nonumber\\
    &-\frac{i\beta}{2}\frac{\tau^4}{12}{}_2F_3\left(1,1;\frac{8}{6},\frac{9}{6},\frac{10}{6};-\left(\frac{\beta\tau^3}{12}\right)^2\right)\nonumber\\
    &-\frac{\alpha^2}{8}\frac{\tau^5}{30}{}_2F_3\left(1,\frac{7}{6};\frac{9}{6},\frac{10}{6},\frac{11}{6};-\left(\frac{\beta\tau^3}{12}\right)^2\right).
\end{align}
\begin{figure}
\begin{center}
\includegraphics[width=0.75\columnwidth]{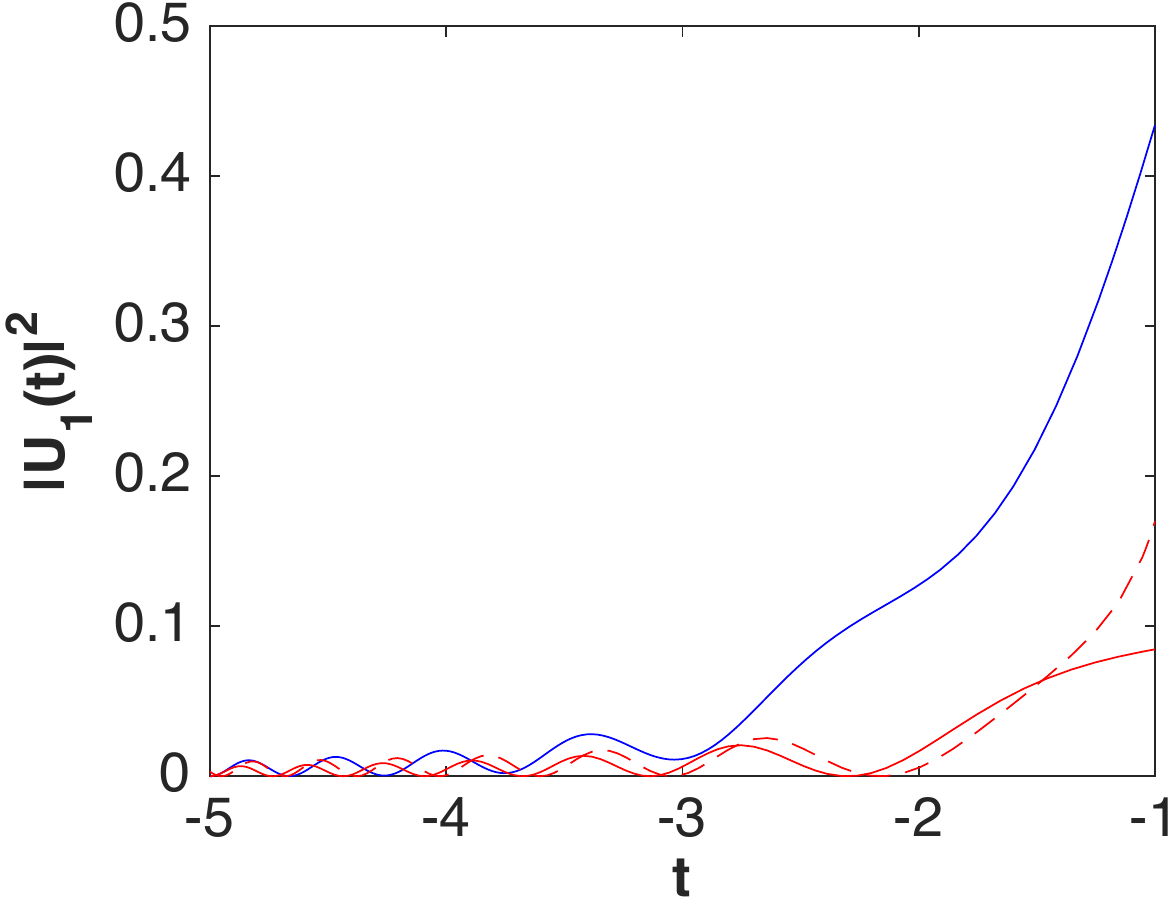}
\end{center}
\caption{The transition probability $|U_1(t)|^2$ subjected to the initial conditions $|U_1(t_i)|=0$ and $|\dot{U}_1(t_i)|=|f|$. Exact solution of Eq.\:\eqref{Tau} is depicted in blue, approximate solution of the homogeneous equation $\bar{U}^{\prime\prime}_1=-\frac{1}{16}\beta^2\tau^4\bar{U}_1$ is depicted in red solid line, and the analytical approximation to Eq.\:\eqref{BesselOneSix}, which is given by Eqs.\:\eqref{HeunBesselAnalytic} and \eqref{HeunBesselLargeT}, is depicted in red dashed line. Here, $t_i=-5$, $t_f=-1$, $|f|=1$, $\alpha=1$, and $\beta=2$.}
\label{fig:HeunBessel}
\end{figure}Here ${}_pF_q(a_1,\cdots,a_p;b_1,\cdots,b_q;z)$ is the generalized hypergeometric function of order $p$, $q$, and $G(\tau)$ has the following asymptotic expansion in the limit $\tau\rightarrow\infty$
\begin{equation}\label{HeunBesselLargeT}
G(\tau)\approx \frac{\alpha^2}{\beta^2\tau}.
\end{equation}
As we can see from Fig.\:\ref{fig:HeunBessel}, the analytical approximation of the transition probability $|U_1|^2$, which is determined by Eq.\:\eqref{HeunBesselAnalytic} with $G(\tau)\approx \alpha^2/\beta^2\tau^{-1}$, agrees with the exact result in the long-time regime. 

\subsection{Analytical Approximations to the Connection Problem}\label{IV}
\begin{figure*}
	\subfloat[$t_i=-10$, $t_f=10$, $c_1^{(III)}=-0.2427+0.7182i$, $c_2^{(III)}=0.4995-0.8049i$\label{sfig:HeunWithBessel}]{%
	\includegraphics[width=\columnwidth]{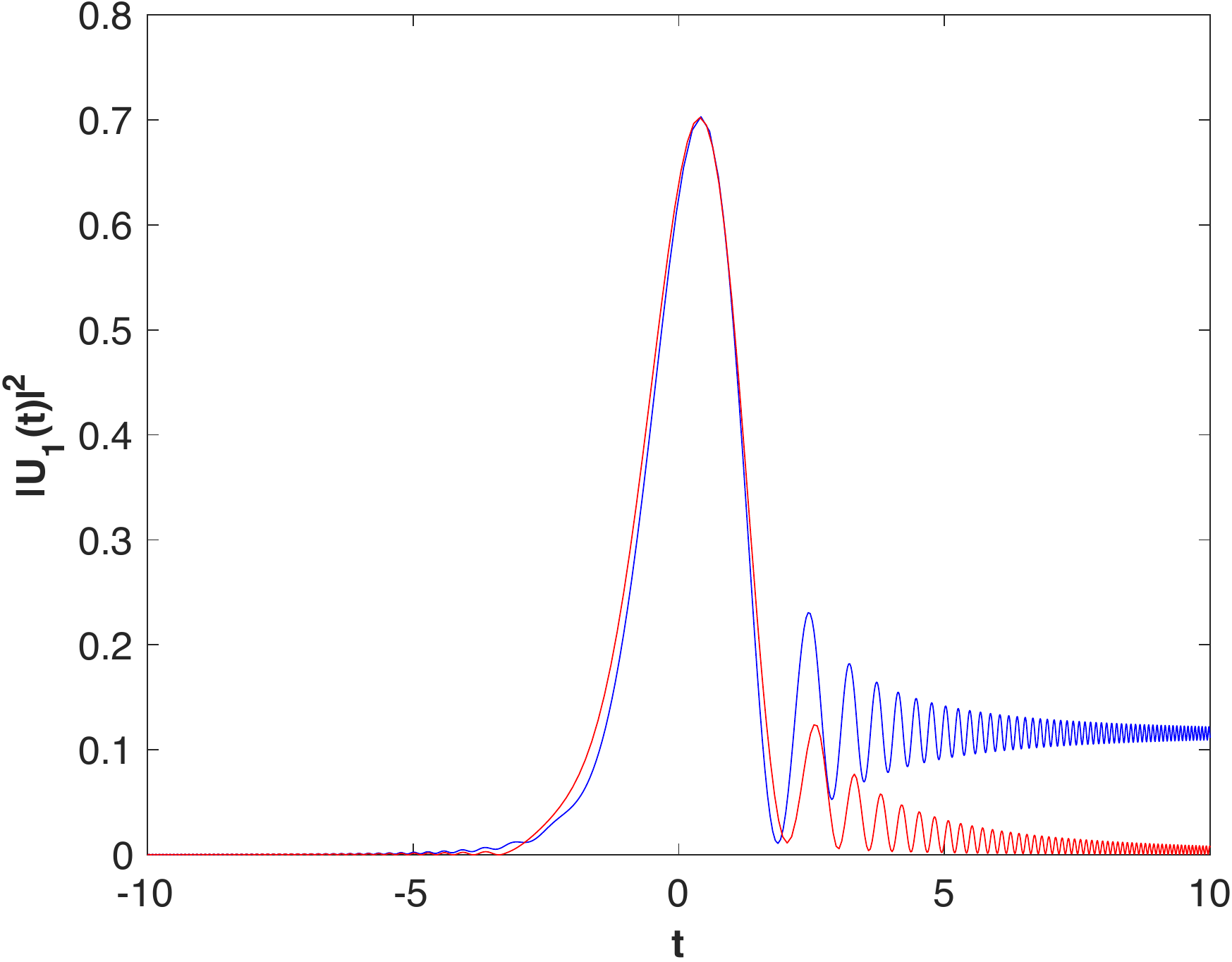}%
	}\hfill
	\subfloat[$t_i=-10$, $t_f=50$, $\rho_f^2=0.1142$, $S_0=-1.0083$, $c_1^{(III)}=-0.6341$, $c_2^{(III)}=0.5761$\label{sfig:HeunWithRho}]{%
 	 \includegraphics[width=\columnwidth]{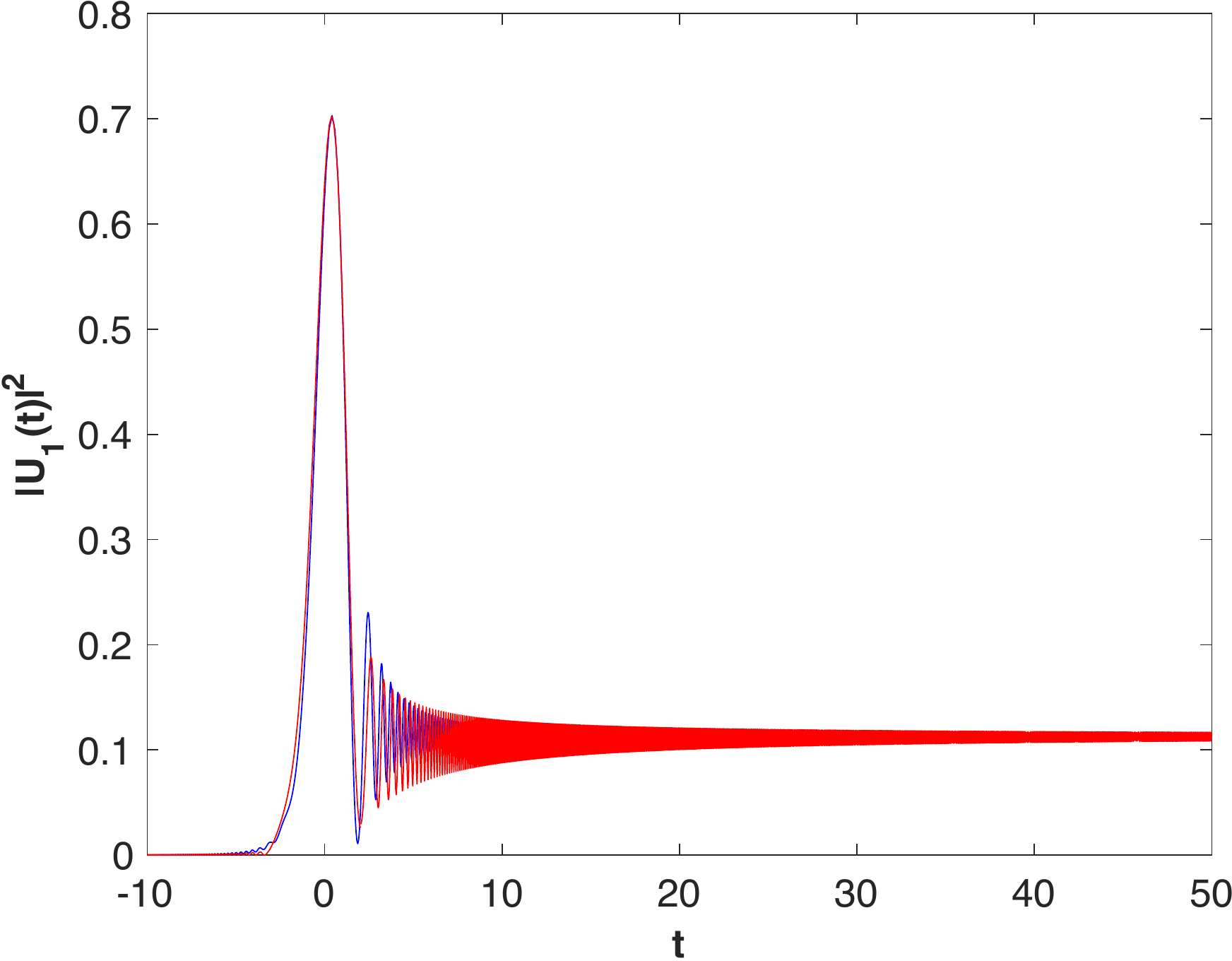}%
	}
\caption{The transition probability $|U_1(t)|^2$ subjected to the initial conditions $U_1(t_i)=0$ and $\dot{U}_1(t_i)=|f|$. Exact solution of Eq.\:\eqref{HeunAlphaZero} is depicted in blue, and the analytical approximations of the transition probability are depicted in red solid lines. In Fig.\:\ref{sfig:HeunWithBessel}, $U_1^{(III)}=c_1^{(III)}\sqrt{t}J_{1/6}(\beta t^3/12)+c_2^{(III)}\sqrt{t}J_{-1/6}(\beta t^3/12)$; whereas in Fig.\:\ref{sfig:HeunWithRho}, $U_1^{(III)}= \rho e^{iS}+c_1^{(III)}\sqrt{t}J_{1/6}(\beta t^3/12)+c_2^{(III)}\sqrt{t}J_{-1/6}(\beta t^3/12)$. Here, $|f|=1$, $\alpha=0$, $\beta=2$, $t_1=-3.2097$, $t_2=2.0335$, and $t^*=(2/\lambda)^{\frac{1}{3}}=1.5874$.}
\label{fig:HeunCompare}
\end{figure*}In the last subsections, we have studied the dynamics for the transition amplitude in both short- and long-time regimes. In this subsection, we focus on the problem of connecting different local solutions of the transition dynamics. As we showed in Fig.\:\ref{sfig:ParabolicCylinderWithoutCorrection}, the transition dynamics for large $\alpha/\beta$ can be well-described by the Landau-Zener formula. Hence, we concentrate on the transition dynamics for small $\alpha/\beta$. Without loss of generality, we only consider the parabolic glancing case with $\alpha=0$. Then Eq.\:\eqref{Tau} may be written as
\begin{equation}\label{HeunAlphaZero}
    \frac{d^2U_1}{dt^2}+\left(|f|^2-2i\lambda t+\lambda^2t^4\right)U_1=0,
\end{equation}
where $\lambda\equiv \beta/4$. Substitution of $U_1\equiv \rho e^{i S}$ into Eq.\:\eqref{HeunAlphaZero} yields the following set of differential equations
\begin{subequations}
\begin{gather}\label{Action1}
    \frac{1}{\rho}\frac{d^2\rho}{dt^2}-\left(\frac{dS}{dt}\right)^2+|f|^2+\lambda^2t^4=0,\\
    \frac{d}{dt}\left(\rho^2\frac{dS}{dt}\right)-2\lambda t\rho^2=0,\label{Action2}
\end{gather}
\end{subequations}
where the initial conditions are $\rho(-\infty)=0$ and $\dot{\rho}(-\infty)=1$. In the long time limit, we expect that the transition probability $\rho^2=|U_1|^2$ converges toward a stationary value. Hence, we assume $\ddot{\rho}\approx 0$ and $\dot{S}^2=|f|^2+\lambda^2t^4$ in Eq.\:\eqref{Action1}, which yields $S(t)\approx \lambda t^3/3+S_0$ and $\frac{d}{dt}\left(\rho^2\sqrt{|f|^2+\lambda^2t^4}\right)=2\lambda t\rho^2$. A direct computation yields
\begin{equation}\label{HeunRho}
\rho(t)\approx \frac{\rho_f}{\sqrt{2}}\sqrt{1+\frac{\lambda t^2}{\sqrt{|f|^2+\lambda^2t^4}}}\equiv\rho_fF(t).
\end{equation}
where $\rho\approx\rho_f(1-\frac{|f|^2}{2\lambda^2}t^{-4})$ and $\dot{\rho}\approx -\frac{2|f|^2}{\lambda^2}\rho_ft^{-5}$ in the long-time limit. We will use this solution in later discussions. 

As we have discussed in the last subsections, the transition dynamics for the parabolic-glancing case can be well-approximated by the Airy and Bessel functions in the short- and long-time regimes. Hence, we only need to connect the local solutions at two critical times $t_1$ and $t_2$ --- for $t\in (-\infty,t_1]$, we have $U_1^{(I)}= c_1^{(I)}w_1(t)+c_2^{(I)}w_2(t)$; for $t\in[t_1,t_2]$, we have $U_1^{(II)}= c_1^{(II)}\Ai(z)+c_2^{(II)}\Bi(z)$ with $z\equiv(\frac{i\beta}{2})^{1/3}(t+\frac{2i}{\beta}|f|^2)$; and for $t\in[t_2,\infty)$, we have $U_1^{(III)}=c_1^{(III)}w_1(t)+c_2^{(III)}w_2(t)$, where $w_1\equiv\sqrt{t}J_{1/6}(\beta t^3/12)$ and $w_2\equiv\sqrt{t}J_{-1/6}(\beta t^3/12)$. Then the connection formulas for the transition amplitudes are
\begin{subequations}
\begin{gather}
U_1^{(I)}=U_1^{(II)},\:\dot{U}_1^{(I)}=\dot{U}_1^{(II)},\:\mbox{for}\:t=t_1;\\
U_1^{(II)}=U_1^{(III)},\:\dot{U}_1^{(II)}=\dot{U}_1^{(III)},\:\mbox{for}\:t=t_2.
\end{gather}
\end{subequations}
A direct computation yields 
\begin{subequations}
\begin{align}
c_1^{(I)}&=-\frac{|f|w_2(t_i)}{\mathscr{W}\{w_1(t_i),w_2(t_i)\}},c_2^{(I)}=\frac{|f|w_1(t_i)}{\mathscr{W}\{w_1(t_i),w_2(t_i)\}},\\
c_1^{(II)}&=\frac{U_1^{(I)}(t_1)\frac{d}{dt}\Bi(z(t_1))-\dot{U}_1^{(I)}(t_1)\Bi(z(t_1))}{\mathscr{W}\{\Ai(z(t_1)),\Bi(z(t_1))\}},\\
c_2^{(II)}&=\frac{\dot{U}_1^{(I)}(t_1)\Ai(z(t_1))-U_1^{(I)}(t_1)\frac{d}{dt}\Ai(z(t_1))}{\mathscr{W}\{\Ai(z(t_1)),\Bi(z(t_1))\}},\\
c_1^{(III)}&=\frac{U_1^{(II)}(t_2)\dot{w}_2(t_2)-\dot{U}_1^{(II)}(t_2)w_2(t_2)}{\mathscr{W}\{w_1(t_2),w_2(t_2)\}},\\
c_2^{(III)}&=\frac{\dot{U}_1^{(II)}(t_2)w_1(t_2)-U_1^{(II)}(t_2)\dot{w}_1(t_2)}{\mathscr{W}\{w_1(t_2),w_2(t_2)\}},
\end{align}
\end{subequations}
where $\mathscr{W}\{\Ai(z(t)),\Bi(z(t))\}=(i\beta/2)^{1/3}/\pi$ is the Wronskian of $\Ai(z(t))$ and $\Bi(z(t))$, $\dot{w}_1(t)=\lambda t^{5/2}J_{-5/6}(\lambda t^3/3)$, $\dot{w}_2(t)=\lambda t^{5/2}J_{5/6}(\lambda t^3/3)$, and $\mathscr{W}\{w_1(t),w_2(t)\}=-3/\pi$ is the Wronskian of $w_1(t)$ and $w_2(t)$. Hence, the dynamics of the transition amplitude is determined up to two critical times $t_1$ and $t_2$. As we can see from Eq.\:\eqref{HeunAlphaZero}, the long-time limit is determined by $\lambda^2t^4\gg 2\lambda t$, which yields $t\gg (2/\lambda)^{1/3}$; whereas the short-time limit is determined by $\lambda^2t^4\ll 2\lambda t$, which yields $t\ll (2/\lambda)^{1/3}$. Hence, we may approximate the critical times $t_1$ and $t_2$ by $-(2/\lambda)^{1/3}$ and $(2/\lambda)^{1/3}$ respectively. In Fig.\:\ref{sfig:HeunWithBessel}, we compare the analytical approximation $|U_1|^2=|U_1^{(I)}|^2\cup |U_1^{(II)}|^2\cup |U_1^{(III)}|^2$ of the transition probability with the numerical solution of Eq.\:\eqref{HeunAlphaZero}. The result shows that the transition probability is well-described by $|U_1^{(I)}|^2\cup |U_1^{(II)}|^2$ for $t\in(-\infty,t_2)$. But for $t\in[t_2,\infty)$, the transition probability, which approaches a non-zero limiting value in the long-time limit, may not be well-described by $|U_1^{(III)}|^2$, as the Bessel function approaches zero at $t\rightarrow \infty$, \textit{e.g.}, $w_1(t)=\sqrt{t}J_{1/6}(\frac{\beta}{12}t^3)\approx \sqrt{\frac{6}{\pi\lambda}}t^{-1}\cos(\frac{\beta}{12}t^3-\frac{\pi}{3})$. In order to fix this problem, we modify the transition amplitude $U_1^{(III)}$ by replacing $c_1^{(III)}w_1+c_2^{(III)}w_2$ by $\rho e^{iS}+c_1^{(III)}w_1+c_2^{(III)}w_2$, where $S\approx \lambda t^3/3+S_0$ and $\rho$ is determined by Eq.\:\eqref{HeunRho}. A direct computation yields
\begin{subequations}
\begin{gather}\label{HeunConnection1}
\rho_f=\frac{\Im{U_1^{(II)}(t_2)}}{F(t_2)}\sqrt{1+\frac{1}{\dot{S}^2(t_2)}\left(\frac{\Im{\dot{U}_1^{(II)}(t_2)}}{\Im{U_1^{(II)}(t_2)}}-\frac{\dot{F}(t_2)}{F(t_2)}\right)^2},\\
S_0=\cot^{-1}\left[\frac{1}{\dot{S}(t_2)}\left(\frac{\Im{\dot{U}_1^{(II)}(t_2)}}{\Im{U_1^{(II)}(t_2)}}-\frac{\dot{F}(t_2)}{F(t_2)}\right)\right]-\frac{\lambda t_2^3}{3},\\
c_1^{(III)}=\frac{\tilde{U}_1^{(II)}(t_2)\dot{w}_2(t_2)-\dot{\tilde{U}}_1^{(II)}(t_2)w_2(t_2)}{\mathscr{W}\{w_1(t_2),w_2(t_2)\}},\\
c_2^{(III)}=\frac{\dot{\tilde{U}}_1^{(II)}(t_2)w_1(t_2)-\tilde{U}_1^{(II)}(t_2)\dot{w}_1(t_2)}{\mathscr{W}\{w_1(t_2),w_2(t_2)\}},
\end{gather}
\end{subequations}
where $\dot{S}^2=|f|^2+\lambda^2t^4$ and $\tilde{U}_1^{(II)}\equiv U_1^{(II)}-\rho_fF\cos(\lambda t^3/3+S_0)$. For $t\rightarrow \infty$, as $w_1$ and $w_2$ decreases as $t^{-1}$, the final transition probability $|U_1(\infty)|^2$ is given by $\rho_f^2$, which is determined by Eq.\:\eqref{HeunConnection1}. In Fig.\:\ref{sfig:HeunWithRho}, we compare the modified analytical approximation of the transition probability with the numerical solution of Eq.\:\eqref{HeunAlphaZero}. The result shows that the transition probability $|U_1|^2$ is well-described by $|U_1^{(I)}|^2\cup |U_1^{(II)}|^2\cup |U_1^{(III)}|^2$ in the whole time range after the replacement $c_1^{(III)}w_1+c_2^{(III)}w_2\rightarrow \rho e^{iS}+c_1^{(III)}w_1+c_2^{(III)}w_2$.

\section{Conclusion}\label{V}
To summarize, we studied the transition dynamics of the parabolic model --- a two-state system subject to a quadratically detuning over an infinite time interval. The solutions are expressed in terms of the tri-confluent Heun functions, which are the generalizations of the conventional confluent hypergeometric functions. Instead of rigorously solving the Stokes multipliers which connect the asymptotic solutions of the transition amplitudes, we derived concise analytical approximations to the transition amplitudes in both short- and long-time regimes, and provided practical formulas for connecting local solutions in different regimes. We gave applicable estimation of the critical times that separate different time regimes. The transition dynamics is shown to be well-described by the analytical formulas in the whole-time range by comparison with exact results.

In future works, we would like to extend our study to super-linear model, in which the laser detuning is a polynomial in time with cubic or higher degrees, and the sub-linear model in which the laser detuning is a rational function in time. We also want to determine the critical times that separate the short- and long-time regimes rigorously, instead of only giving a rough estimate of the magnitude.
 
\begin{acknowledgements}
The Authors would like to thank the Science and Technology Development Fund of the Macau SAR for providing support, FDCT 023/2017/A1.
\end{acknowledgements}

\begin{appendix}

\section{Equations of Motion for a Two-level Atom Dipole-Interacting with a Classical Driving Field}\label{A}
In this appendix, we provide the background knowledge of the two-level atom description of light-matter integrations for readers. When the two target atomic levels are nearly resonant with the driving field, while on the same time the other atomic levels are detuned far off resonance, we may regard the system as two discrete non-degenerate states, \textit{e.g.}, the ground state $|g\rangle$ and the first excited state $|e\rangle$, then the Hamiltonian of the atom may be written as $\hat{H}_{atom}=E_e|e\rangle\langle e|+E_g|g\rangle\langle g|$. For the case when the atom interacts with an external electric field under dipole approximation, the interaction Hamiltonian becomes $\hat{H}_{int}=-\hat{\mathbf{d}}\cdot\mathbf{E}(\hat{\mathbf{r}}_{cm})$, where $\hat{\mathbf{r}}_{cm}$ is the position operator for the center of mass. When the De Broglie wavelength of the atom is small compared to the interatomic spacing, the center of mass position may be treated classically \cite{garrison2008quantum}. 

In the following, we discuss the coherent excitation of the two-level atom under the driving of an external electric field. Let us denote the state of the two-level atom as $|\psi(t)\rangle=c_e(t)|e\rangle+c_g(t)|g\rangle$, the Schr\"{o}dinger equations for the two wave amplitudes have the form
\begin{subequations}
\begin{align}
    i\hbar\frac{dc_e}{dt}&=(E_e(t)-\langle e|\hat{\mathbf{d}}\cdot\mathbf{E}|e\rangle)c_e-\langle e|\hat{\mathbf{d}}\cdot\mathbf{E}|g\rangle c_g,\\
    i\hbar\frac{dc_g}{dt}&=-\langle g|\hat{\mathbf{d}}\cdot\mathbf{E}|e\rangle c_e+(E_g(t)-\langle g|\hat{\mathbf{d}}\cdot\mathbf{E}|g\rangle)c_g,
\end{align}
\end{subequations}
where $\hat{\mathbf{d}}\equiv q\hat{\mathbf{x}}$ is the transition electric dipole moment of the atom. For atom that possesses inversion symmetry, the energy eigenstates are symmetric or anti-symmetric, and hence the expectation values of the dipole moment vanishes, $\langle e|\hat{\mathbf{d}}|e\rangle=\langle g|\hat{\mathbf{d}}|g\rangle=0$. Let us denote the energies of the two atomic levels as $E_e=\frac{\hbar}{2}\omega_0$ and $E_g=-\frac{\hbar}{2}\omega_0$, where $\hbar\omega_0\equiv E_e-E_g$ is the energy difference between the two atomic levels. For a monochromatic wave, the electric field may be written as $\mathbf{E}=\frac{1}{2}\{E_0\exp[i\int_0^t\omega(\tau)d\tau]\mathbf{e}_p+h.c.\}$, where $E_0$ is the complex electric field amplitude at the position $\mathbf{r}_{cm}$, and $\mathbf{e}_p$ is the polarization vector of the incident electric field. If we denote the transition dipole moment $\langle e|\hat{\mathbf{d}}|g\rangle$ as $\mathbf{d}_{ge}$, the Schr\"{o}dinger equations for the two wave amplitudes may be simplified as
\begin{subequations}
\begin{align}\label{TwoLevel1}
    i\frac{dc_e}{dt}&=\frac{\omega_0}{2}c_e-\frac{\mathbf{d}_{ge}}{2\hbar}\cdot\left[E_0\mathbf{e}_pe^{i\int_0^t\omega(\tau)d\tau}+h.c.\right]c_g,\\
    i\frac{dc_g}{dt}&=-\frac{\mathbf{d}^*_{ge}}{2\hbar}\cdot\left[E_0\mathbf{e}_pe^{i\int_0^t\omega(\tau)d\tau}+h.c.\right]c_e-\frac{\omega_0}{2}c_g.\label{TwoLevel2}
\end{align}
\end{subequations}
In the reference frame rotating about the $z$-axis with frequency $\omega(t)$, the state of the two-level atom is $|\psi'(t)\rangle=\exp(\frac{i}{\hbar}\int_0^\tau\omega(\tau)dtS_z)|\psi(t)\rangle$. Then the relationships between the wave amplitudes in the rotated and unrotated frame are
\begin{equation}\label{RWA}
a_e=e^{\frac{i}{2}\int_0^t\omega(\tau)d\tau}c_e, a_g=e^{-\frac{i}{2}\int_0^t\omega(\tau)d\tau}c_g.
\end{equation}
Substitution of Eq.\:\eqref{RWA} into Eqs.\:\eqref{TwoLevel1} - \eqref{TwoLevel2} yields
\begin{align*}
    i\frac{da_e}{dt}&= -\frac{\Delta}{2}a_e- \left[\frac{E_0}{2\hbar}\mathbf{d}_{ge}\cdot\mathbf{e}_pe^{2i\int_0^t\omega(\tau)d\tau}+\frac{E_0^*}{2\hbar}\mathbf{d}_{ge}\cdot\mathbf{e}_p^*\right]a_g,\\
    i\frac{da_g}{dt}&=-\left[\frac{E_0}{2\hbar}\mathbf{d}^*_{ge}\cdot\mathbf{e}_p+\frac{E_0^*}{2\hbar}\mathbf{d}^*_{ge}\cdot\mathbf{e}_p^*e^{-2i\int_0^t\omega(\tau)d\tau}\right]a_e+\frac{\Delta}{2}a_g,
\end{align*}
where $\Delta\equiv \omega-\omega_0$ is the laser detuning. Applying the rotating wave approximation, we neglect the fast-oscillating terms like $\exp\{\pm2i\int_0^t\omega(\tau)d\tau\}$ in the Schr\"{o}dinger equations, and obtain
\begin{equation}
    i\frac{da_e}{dt}= -\frac{\Delta}{2}a_e +fa_g, i\frac{da_g}{dt}=f^*a_e+\frac{\Delta}{2}a_g,
\end{equation}
where $f\equiv -\frac{E_0^*}{2\hbar}\mathbf{d}_{ge}\cdot\mathbf{e}_p^*$ is the Rabi frequency for the transition dipole moment. For the special case that $\omega=\omega_0$, the laser detuning $\Delta$ vanishes, and the coupled-mode equations become $i\dot{a}_e=fa_g$ and $i\dot{a}_g=f^*a_e$, which are equivalent to
\begin{equation}\label{RabiOde}
    \ddot{a}_e-\frac{\dot{f}}{f}\dot{a}_e+|f|^2a_e=0, \ddot{a}_g-\frac{\dot{f}^*}{f^*}\dot{a}_g+|f|^2a_g=0.
\end{equation}
When the atom is initially at the ground state, we have $c_g(0)=1$ and $c_e(0)=0$, or equivalently $a_g(0)=1$ and $a_e(0)=0$. For the special case that the Rabi frequency is time-independent, Eq.\:\eqref{RabiOde} is solved by $a_e=\sin ft$ and $a_g=\cos ft$, or equivalently $c_e=e^{-\frac{i}{2}\int_0^t \omega(\tau)d\tau}\sin ft$ and $c_g=e^{\frac{i}{2}\int_0^t \omega(\tau)d\tau}\cos ft$ in the lab frame, which results in the occupation probabilities $|c_e|^2=\sin^2 ft$ and $|c_g|^2=\cos^2 ft$.

\section{The Landau-Zener Model}\label{B}
In this appendix, we provide supplementary information on the large variable expansion of the transition probability in the Landau-Zener model. In the Landau-Zener model, the wave amplitude after the change of variable $U_1=C_1\exp\{\frac{i\alpha}{2}\int_0^t sds\}$ is governed by the Schr\"{o}dinger equation $\ddot{U}_1+J(t)U_1=0$, where $J(t)=|f|^2-\frac{i\alpha}{2}+\frac{\alpha^2t^2}{4}$. After the change of variable $z=e^{-i\pi/4}\alpha^{1/2}t$ and $n=i|f|^2/\alpha$, the Schr\"{o}dinger equation becomes the parabolic cylinder equation $U_1^{\prime\prime}+(n+\frac{1}{2}-\frac{1}{4}z^2)U_1=0$, and is solved by the parabolic cylinder function $D_n(z)$, which has the following asymptotic expansion for large $|z|$
\begin{subequations}
\begin{align}\label{AsymptoticExpansion1}
    D_n(z)&\approx e^{-\frac{1}{4}z^2}z^n\sum_{k=0}^\infty\frac{(-n)_{2k}}{k!(-2z^2)^k},\:|\arg z|<\frac{3\pi}{4},\\
    D_n(z)&\approx e^{-\frac{1}{4}z^2}z^n\sum_{k=0}^\infty\frac{(-n)_{2k}}{k!(-2z^2)^k}-\frac{\sqrt{2\pi}e^{\pm i\pi n}}{\Gamma(-n)}\frac{e^{\frac{1}{4}z^2}}{z^{n+1}}\sum_{k=0}^\infty\frac{(n+1)_{2k}}{k!(2z^2)^k}\nonumber\\
    &\approx e^{-\frac{1}{4}z^2}z^n\left\{1-\frac{n(n-1)}{2z^2}+\cdots\right\}-\frac{\sqrt{2\pi}e^{\pm i\pi n}}{\Gamma(-n)}\frac{e^{\frac{1}{4}z^2}}{z^{n+1}}\nonumber\\
    &\left\{1+\frac{(n+1)(n+2)}{2z^2}+\cdots\right\},\:|\arg z|\in(\frac{\pi}{4},\frac{5\pi}{4}).
\end{align}
\end{subequations}
For $\alpha>0$, we have $z=e^{-i\pi/4}\sqrt{|\alpha|}t$, whereas for $\alpha<0$, we have $z=e^{i\pi/4}\sqrt{|\alpha|}t$. For $t\rightarrow-\infty$, we have $z=e^{\pm i3\pi/4}R$ with $R\equiv \sqrt{|\alpha|}|t|$, which yields
\begin{subequations}
\begin{align*}
    C_1(t\rightarrow-\infty)&=A_{\pm}e^{-i\alpha t^2/4}D_{i\gamma}(e^{\pm3i\pi/4}R)\approx A_{\pm}e^{\mp3\pi\gamma/4}R^{i\gamma},\\
    C_2(t\rightarrow-\infty)&=\frac{ie^{\frac{i}{2}\alpha t^2}}{f}\dot{C}_1\approx\frac{-\gamma e^{\frac{i}{2}\alpha t^2}}{f t}C_1,\:(\gamma\equiv |f|^2/\alpha).
\end{align*} 
\end{subequations}
Hence, $|C_1(-\infty)|=|A_{\pm}|e^{\mp 3\pi\gamma/2}$ and $|C_2(-\infty)|=0$, which violates the initial conditions $C_1(-\infty)=0$ and $|C_2(-\infty)|=1$. We now check the linearly independent solutions $D_{-n-1}(\mp iz)=D_{-n-1}(e^{\pm i\pi/4}R)$. From Eq.\:\eqref{AsymptoticExpansion1}, we obtain
\begin{subequations}
\begin{align}
    C_1(t\rightarrow-\infty)&\approx A_{\pm}e^{\pm \pi( \gamma-i)/4}e^{-i\alpha t^2/2}R^{-i\gamma-1},\\
    C_2(t\rightarrow-\infty)&\approx \frac{\alpha t e^{i\alpha t^2/2}}{f}C_1,
\end{align}
\end{subequations}
which yields $|C_1(-\infty)|=0$ and $|C_2(-\infty)|=|A_{\pm}||\gamma|^{-1/2}e^{|\gamma|\pi/4}$. Hence, the choice of parameters $A_{\pm}=|\gamma|^{1/2}e^{-|\gamma|\pi/4}$ satisfies the initial conditions $C_1(-\infty)=0$ and $|C_2(-\infty)|=1$. For $t\rightarrow+\infty$, we have $\mp iz=e^{\mp3i\pi/4}R$, which yields
\begin{subequations}
\begin{align}
    C_1(t\rightarrow\infty)&\approx \frac{\sqrt{2\pi}|\gamma|^{1/2}e^{-\pi|\gamma|/2}}{\Gamma(i\gamma+1)}R^{i\gamma},\\
    C_2(t\rightarrow\infty)&\approx\frac{ie^{i\alpha t^2/2}}{f}\left(|\gamma|^{1/2}\sqrt{|\alpha|}e^{-\pi|\gamma|}e^{\pm3\pi i/4}R^{-i\gamma}\right).
\end{align}
\end{subequations}
A direct computation yields $|C_2(\infty)|^2=e^{-2\pi|\gamma|}$ and $|C_1(\infty)|^2=1-e^{-2\pi|\gamma|}$.

\section{Large variable solution of $y^{\prime\prime}\pm i\Delta y^\prime+|f|^2y=0$}\label{C}
In this appendix, we consider a slightly general form of Eqs.\:\eqref{TwoLevelProblem2} and \eqref{TwoLevelProblem1}, and analyze the large variable solution of $y^{\prime\prime}\pm i\Delta y^\prime+|f|^2y=0$, where $\Delta\equiv\alpha t+\frac{1}{2}\beta t^2+\cdots$ is a polynomial of time. Let us denote $y=UV$, where $V$ obeys $\pm i\Delta\dot{V}+|f|^2V=0$, and is solved by $V=V_0\exp\{\mp i|f|^2\int_{t_0}^t \Delta^{-1}dt'\}$. Substitution of $y=UV$ into $y^{\prime\prime}\pm i\Delta y^\prime+|f|^2y=0$ yields
\begin{equation}\label{SuperHeunU}
    \ddot{U}-\left(\frac{2i|f|^2}{\Delta}\mp i\Delta\right)\dot{U}-\left(\frac{|f|^4}{\Delta^2}\pm\frac{i|f|^2\dot{\Delta}}{\Delta^2}\right)U=0.
\end{equation}
For large $|t|$, Eq.\:\eqref{SuperHeunU} is approximated by $\ddot{U}\pm i\nu\dot{U}=0$, which is solved by $U\approx U_1\int_{t_0}^t\exp\{\mp i\int_{t_0}^{t'}\Delta dt''\}dt'+U_0$. For the Landau-Zener case that $\Delta=\alpha t$, we obtain $V=V_0\left(t/t_0\right)^{\mp i\gamma}$, $U\approx U_1e^{\pm i\alpha t_0^2/2} \sqrt{\frac{2}{\pm i\alpha}}\int_{z_0}^ze^{-z'^2}dz'+U_0$, where $\gamma\equiv |f|^2/\alpha$ and $z\equiv \sqrt{\pm \frac{i\alpha}{2}}t$. Substitution of the result into $y=UV$ yields
\begin{gather*}
    y\approx c_1\sqrt{\frac{\pi}{\pm 2i\alpha}}\erfc\left(\sqrt{\frac{\pm i\alpha}{2}}t\right)t^{\mp i\gamma}+c_2t^{\mp i\gamma},\nonumber\\
    c_1=-V_0t_0^{\pm i\gamma}U_1e^{\pm i\alpha t_0^2/2},c_2=V_0t_0^{\pm i\gamma}U_0-c_1\sqrt{\frac{\pi}{\pm 2i\alpha}}\erfc(z_0),
\end{gather*}
where the complementary error function $\erfc(z)$ has the following asymptotic expansion
\begin{subequations}
\begin{align}
    \erfc(z)&=\frac{e^{-z^2}}{\sqrt{\pi}z}\sum_{n=0}^{n-1}(-1)^k\frac{(2k-1)!!}{(2z^2)^k}+R_n(z),\\
    R_n(z)&\equiv\frac{(-1)^n}{\sqrt{\pi}}
\frac{(2n)!}{2^{2n-1}n!}\int_z^\infty s^{-2n}e^{-s^2}ds.
\end{align}
\end{subequations}
Hence, for large $|t|$, we obtain
\begin{equation}
    y(t)\approx \frac{c_1}{\pm i\alpha}e^{\mp i\alpha t^2/2}t^{\mp i\gamma-1}+ c_2t^{\mp i\gamma}.
\end{equation}
As a result, the wave amplitudes $C_1$ and $C_2$ in Eqs.\:\eqref{TwoLevelProblem2} and \eqref{TwoLevelProblem1} subjected to the conditions $C_1(-\infty)=0$ and $|C_2(-\infty)|=1$ are asymptotically determined by
\begin{equation}
    C_1(t)\approx \frac{c_1}{i\alpha}e^{-i\alpha t^2/2}t^{-i\gamma-1},
    C_2(t)\approx\frac{-ic_1}{f}t^{-i\gamma}.
\end{equation}
Hence, we obtain $|C_2(\infty)|/|C_2(-\infty)|=e^{-\pi\gamma}$. We now analyze the parabolic model with $\Delta=\alpha t +\frac{1}{2}\beta t^2$. Following similar procedures described previously, we obtain
\begin{subequations}
\begin{gather}
    V=V_0\left(\frac{t}{t+2\alpha/\beta}\right)^{\mp i\gamma},\\
    U\approx U_2\int_{t_0}^t\exp\left\{\mp i\left(\frac{\alpha}{2}t'^2+\frac{\beta}{6}t'^3\right)\right\}dt'+U_0,
\end{gather}
\end{subequations}
where $U_2\equiv U_1\exp\{\pm i(\frac{\alpha}{2}t_0^2+\frac{\beta}{6}t_0^3)\}$. We now analyze the large $|t|$ behavior of the integral $\int_t^\infty e^{-\pi_n(t')}dt'$, where $\pi_n(t)$ is an $n$-order polynomial of $t$, \textit{e.g.}, $\pi_3(t)=\pm i(\frac{\alpha}{2}t^2+\frac{\beta}{6}t^3)$. A direct computation gives
\begin{subequations}
\begin{gather}
    \frac{d}{dt}\left(\frac{e^{-\pi_n}}{\dot{\pi}_n}\right)=-e^{-\pi_n}-\frac{e^{-\pi_n}\ddot{\pi}_n}{\dot{\pi}_n^2},\\
    \frac{d}{dt}\left(\frac{e^{-\pi_n}\ddot{\pi}_n}{\dot{\pi}_n^3}\right)=-\frac{e^{-\pi_n}\ddot{\pi}_n}{\dot{\pi}_n^2}-\frac{3e^{-\pi_n}\ddot{\pi}_n^2}{\dot{\pi}_n^4}+\frac{e^{-\pi_n}\pi_n^{(3)}}{\dot{\pi}_n^3}.
\end{gather}
\end{subequations}
which yields
\begin{equation}
    \int_t^\infty e^{-\pi_n(t')}dt'\approx \frac{e^{-\pi_n(t)}}{\dot{\pi}_n(t)}\left[1-\frac{\ddot{\pi}(t)}{\dot{\pi}_n^2(t)}+\cdots\right].
\end{equation}
Hence, we obtain
\begin{equation}
    y(t)\approx c_1\frac{e^{\mp i(\frac{\alpha}{2}t^2+\frac{\beta}{6}t^3)}}{\pm i(\alpha t+\frac{\beta}{2}t^2)}\left(\frac{t}{t+2\alpha/\beta}\right)^{\mp i\gamma}+c_2\left(\frac{t}{t+2\alpha/\beta}\right)^{\mp i\gamma},
\end{equation}
where $c_1=-V_0U_2$ and $c_2=V_0\{U_0+U_2\int_{t_0}^\infty\exp[\mp i(\frac{\alpha}{2}t^2+\frac{\beta}{6}t^3)]dt\}$. Hence, we obtain
\begin{subequations}
\begin{align}
    C_1(t)&\approx c_1\frac{e^{-i(\frac{\alpha}{2}t^2+\frac{\beta}{6}t^3)}}{i(\alpha t+\frac{\beta}{2}t^2)}\left(\frac{t}{t+2\alpha/\beta}\right)^{-i\gamma},\\
    C_2(t)&\approx \frac{-ic_1}{f}\left(\frac{t}{t+2\alpha/\beta}\right)^{-i\gamma},
\end{align}
\end{subequations}
where $|C_2(\infty)|/|C_2(-\infty)|=1$, which is not consistent with the exact solution of the final transition probability. Hence, we conclude that the asymptotic solutions alone cannot guarantee a correct final transition probability, and a more careful treatment of the connection problem is needed for the parabolic model.

\section{Integrals involving products of Airy, parabolic cylinder, and Bessel functions}\label{AiryAppendix}
To begin with, let us derive some general results regarding integrals of the form $I_n\equiv \int z^ny^2dz$ and $J_n\equiv \int z^ny_1y_2dz$, where $y_1$ and $y_2$ are solutions of the differential equation $y^{\prime\prime}=fy$. If we write $I_n=Py^2+Qyy^\prime+Ry^{\prime 2}$, we obtain $J_n=Py_1y_2+\frac{1}{2}Q(y_1y_2^\prime+y_2y_1^\prime)+Ry_1^{\prime}y_2^{\prime}$, where $P$ and $Q$ satisfy $P=\frac{1}{2}R^{\prime\prime}-fR$ and $Q=-R^\prime$, and $R$ is a solution of the third-order differential equation $R^{\prime\prime\prime}-4fR^\prime-2f^\prime R=2z^n$. As a consequence, $\mathcal{R}\equiv (n+3)(n+2)(n+1)R/2-z^{n+3}$ is a solution of the third-order differential equation $\mathcal{R}^{\prime\prime\prime}-4f\mathcal{R}^\prime-2f^\prime \mathcal{R}=2z^{n+3}(2(n+3)f/z+f^\prime)$. Let us denote $y_1$ and $y_2$ as two independent solutions of $y^{\prime\prime}=fy$ and $\mathscr{W}\{y_1,y_2\}$ being the Wronskian of $y_1$ and $y_2$. A direct computations shows that $\mathscr{W}$ is a constant. Then for any linear combinations of $y_1$ and $y_2$, \textit{i.e.}, $y\equiv\alpha y_1+\beta y_2$, we have
\begin{align}\label{IntegralOperator}
    \mathscr{L}_ny &\equiv \mathscr{W}^{-1}\left(y_2\int y_1z^nydz-y_1\int y_2z^nydz\right)\nonumber\\
    &= \frac{1}{2}R^\prime y-Ry^\prime.
\end{align}

\subsection{Integrals of products of Airy functions}
In this sub-appendix, we evaluate indefinite integrals of the form $I_n\equiv\int z^n y^2dz$ and $J_n\equiv\int z^n y_1y_2dz$ in terms of Airy functions and their first derivatives, where $y_1$ and $y_2$ are solutions of the Airy differential equation $y^{\prime\prime}=zy$. A direct computation shows the following recursion relation
\begin{subequations}
\begin{align}
    I_n&= \frac{1}{2n+1}\left\{\left[z^{n+1}-\frac{n(n-1)}{2}z^{n-2}\right]y^2+nz^{n-1}yy^\prime-z^ny^{\prime 2}\right\}\nonumber\\
    &+\frac{n(n-1)(n-2)}{2(2n+1)}I_{n-3}\\
    &\equiv\left(\frac{1}{2}R_n^{\prime\prime}-zR_n\right)y^2-R_n^\prime yy^\prime+R_ny^{\prime 2},
\end{align}
\end{subequations}
where $R_n$ satisfies the recursion relation
\begin{subequations}
\begin{gather}
    R_n=\frac{1}{2(2n+1)}\left[n(n-1)(n-2)R_{n-3}-2z^n\right],\\
    R_0=-1, R_1=-z/3,R_2=-z^2/5,
\end{gather}
\end{subequations}
which yields $R_3=-(z^3+3)/7$ and $R_4=-(z^4+4z)/9$. Hence, a direct computation yields
\begin{subequations}
\begin{gather}\label{AiryIntegration1}
    I_0=zy^2-y^{\prime 2}, I_1=\frac{1}{3}\left(z^2y^2+y'y-zy'^2\right),\\
    I_2=\frac{1}{5}\left((z^3-1)y^2+2zy'y-z^2y^{\prime 2}\right),\\
    I_3=\frac{1}{7}\left(z^4y^2+3z^2y'y-(z^3+3)y'^2\right),\\
    I_4=\frac{1}{9}\left((z^5-2z^2)y^2+4(z^3+1)y'y-(z^4+4z)y^{\prime 2}\right).\label{AiryIntegration5}
\end{gather}
\end{subequations}
Let us denote $y_1=\Ai(z)$, $y_2=\Bi(z)$, and $y=\alpha y_1+\beta y_2$. From Eq.\:\eqref{IntegralOperator}, we obtain
\begin{subequations}
\begin{gather}
    \mathscr{L}_0y=y',\mathscr{L}_1y=\frac{z}{3}y'-\frac{1}{6}y,\\
     \mathscr{L}_2y=\frac{z^2}{5}y'-\frac{z}{5}y,\mathscr{L}_3y=\frac{z^3+3}{7}y'-\frac{3z^2}{14}y,\\
    \mathscr{L}_4y=\frac{z^4+4z}{9}y'-\frac{2z^3+2}{9}y.
\end{gather}
\end{subequations}

\subsection{Integrals of products of parabolic cylinder functions}
In this sub-appendix, we evaluate indefinite integrals of the form $I_n\equiv\int z^n y^2dz$ and $J_n\equiv\int z^n y_1y_2dz$ in terms of parabolic cylinder functions and their first derivatives, where $y_1$ and $y_2$ are both solutions of the parabolic cylinder differential equation $y^{\prime\prime}=(a+\frac{1}{4}z^2)y$. If we write $I_n\equiv\left(\frac{1}{2}R_n^{\prime\prime}-zR_n\right)y^2-R_n^\prime yy^\prime+R_ny^{\prime 2}$, we obtain $R_n^{\prime\prime\prime}-(4a+z^2)R_n^\prime-zR=2z^n$. We may expand $R_n$ in a series as $\sum_{k=0}^\infty c_kz^k$, where the coefficients $c_k$ satisfies the following three-term recursion relation
\begin{equation*}
(k+2)(k+1)kc_{k+2}-4akc_k-(k-1)c_{k-2}=\begin{cases}
    2,\:k=n+1;\\
    0,\:\mbox{otherwise}.\\
    \end{cases}
\end{equation*}
A direct computation yields
\begin{subequations}
\begin{align}\label{WeberIntegral1}
    I_1&=\frac{1}{2}(z^2+4a)y^2-2y^{\prime 2},\\
    I_3&=\frac{1}{3}\left[\left(\frac{z^4}{2}-2az^2-16a^2-2\right)y^2\right.\nonumber\\
    &\left.+4zy'y+(16a-2z^2)y^{\prime 2}\right],\label{WeberIntegral2}
\end{align}
\end{subequations}
which corresponds to $R_1=-2$ and $R_3=(16a-2z^2)/3$. Let us denote $y_1=U(a,z)$, $y_2=U(-a,\mp iz)$, and $y=\alpha y_1+\beta y_2$. From Eq.\:\eqref{IntegralOperator}, we obtain
\begin{equation}
    \mathscr{L}_1y=2y^\prime,\\
    \mathscr{L}_3y=-\frac{2z}{3}y-\frac{16a-2z^2}{3}y^\prime.
\end{equation}

\subsection{Integrals of products of Bessel functions}
In this sub-appendix, we evaluate indefinite integrals of the form $I_n\equiv\int \tau^n y^2d\tau$ and $J_n\equiv\int \tau^n y_1y_2d\tau$ in terms of Bessel functions and their first derivatives, where $y_1$ and $y_2$ are both solutions of the differential equation $y^{\prime\prime}=-\lambda^2\tau^4y$. Using the ansatz $I_n=\left(\frac{1}{2}R_n^{\prime\prime}+\lambda^2\tau^4 R_n\right)y^2-R_n^\prime yy^\prime+R_ny^{\prime 2}$, we obtain $R_n^{\prime\prime\prime}+4\lambda^2\tau^4R_n^\prime+8\lambda^2\tau^3R_n=2\tau^n$, where $R_n$ satisfies the following recursion relation
\begin{equation}
    R_n=\frac{2\tau^{n+3}-4\lambda^2(n+5)R_{n+6}}{(n+1)(n+2)(n+3)}.
\end{equation}
A direct computation yields
\begin{subequations}
\begin{gather}
    I_3=\frac{\tau^4}{4}y^2+\frac{1}{4\lambda^2}y^{\prime 2},\\
    I_4=\frac{\tau^5}{6}y^2-\frac{1}{6\lambda^2}yy^\prime+\frac{\tau}{6\lambda^2}y^{\prime 2},\\
    I_5=(\frac{\tau^6}{8}+\frac{1}{8\lambda^2})y^2-\frac{\tau}{4\lambda^2}yy^\prime+\frac{\tau^2}{8\lambda^2}y^{\prime 2},
\end{gather}
\end{subequations}
which corresponds to $R_3=1/(4\lambda^2)$, $R_4=\tau/(6\lambda^2)$ and $R_5=\tau^2/(8\lambda^2)$. We need to evaluate $I_n$ with $n=0, 1, 2$. We may expand $R_n$ in a series as $\sum_{k=0}^\infty c_k\tau^k$, where the coefficients $c_k$ satisfy the following two-term recursion relation
\begin{equation*}
    (k+6)(k+5)(k+4)c_{k+6}+4\lambda^2(k+2)c_k=
    \begin{cases}
    2,\:k=n-3;\\
    0,\:\mbox{otherwise}.\\
    \end{cases}
\end{equation*}
A direct computation yields
\begin{align}
    R_n&=\frac{2\tau^{n+3}}{(n+3)(n+2)(n+1)}\left[1-\frac{4\lambda^2(n+5)\tau^6}{(n+9)(n+8)(n+7)}+\cdots\right]\nonumber\\
    &=\frac{2\tau^{n+3}{}_2F_3\left(1,\frac{n+5}{6};\frac{n+7}{6},\frac{n+8}{6},\frac{n+9}{6};-\left(\frac{\lambda\tau^3}{3}\right)^2\right)}{(n+3)(n+2)(n+1)},
\end{align}
where ${}_pF_q(a_1,\cdots,a_p;b_1,\cdots,b_q;z)$ is the generalized hypergeometric function of order $p$, $q$. To be specific, let us denote $y_1=\sqrt{\tau}J_{1/6}(\lambda\tau^3/3)$, $y_2=\sqrt{\tau}J_{-1/6}(\lambda\tau^3/3)$, and $y=\alpha y_1+\beta y_2$. From Eq.\:\eqref{IntegralOperator}, we obtain ($n=0,1,2$)
\begin{align}
    \mathscr{L}_ny&=\frac{\tau^{n+2}{}_2F_3\left(1,\frac{n+5}{6};\frac{n+3}{6},\frac{n+7}{6},\frac{n+8}{6};-\left(\frac{\lambda\tau^3}{3}\right)^2\right)}{(n+2)(n+1)}y\nonumber\\
    &-\frac{2\tau^{n+3}{}_2F_3\left(1,\frac{n+5}{6};\frac{n+7}{6},\frac{n+8}{6},\frac{n+9}{6};-\left(\frac{\lambda\tau^3}{3}\right)^2\right)}{(n+3)(n+2)(n+1)}y^\prime.
\end{align}
In particular, for $z\rightarrow\infty$ and $p=q-1$, the generalized hypergeometric function ${}_pF_q(\mathbf{a};\mathbf{b};-z)$ has the following asymptotic expansion in $|\arg(z)|<\pi$
\begin{gather}
    \frac{\prod_l\Gamma(a_l)}{\prod_l\Gamma(b_l)}{}_pF_q(\mathbf{a};\mathbf{b};-z)\approx H_{p,q}(z)+E_{p,q}(ze^{-i\pi})+E_{p,q}(ze^{i\pi}),\nonumber\\
    E_{p,q}(z)\approx (2\pi)^{(p-q)/2}\kappa^{-\nu-1/2}e^{\kappa z^{1/\kappa}}(\kappa z^{1/\kappa})^\nu,\nonumber\\
    H_{p,q}(z)\approx \sum_{m=1}^p \Gamma(a_m)\left(\frac{\prod_{l\neq m}\Gamma(a_l-a_m)}{\prod_l\Gamma(b_l-a_m)}\right)z^{-a_m},
\end{gather}
where $\kappa\equiv q-p+1$, $\nu\equiv \sum_la_l-\sum_lb_l +(q-p)/2$. For $R_n$ with $n=0$ or $2$, we have $\kappa=2$, $\nu=-(n+5)/3$, and 
\begin{gather}
    E_{2,3}(ze^{\pm i\pi})\approx \frac{z^{-\frac{5+n}{6}}}{\sqrt{4\pi}}e^{\pm i(2\sqrt{z}-\frac{(5+n)\pi}{6})},\nonumber\\
    H_{2,3}(z)\approx \frac{\Gamma(\frac{n-1}{6})z^{-1}}{\Gamma(\frac{n+1}{6})\Gamma(\frac{n+2}{6})\Gamma(\frac{n+3}{6})}\pm\frac{ 2\pi z^{-\frac{n+5}{6}}}{\Gamma(\frac{2}{6})\Gamma(\frac{3}{6})\Gamma(\frac{4}{6})},
\end{gather}
where the plus and minus signs are for $n=0$ and $n=2$ respectively. For $n=0$, we have $z^{-1}\ll z^{-5/6}$; whereas for $n=2$, we have $z^{-7/6}\ll z^{-1}$. Hence, a direct computation yields $R_2(\tau)\approx\frac{8}{\beta^2\tau}$ and
\begin{equation*}
    R_0(\tau)\approx\frac{\sqrt{3}}{6^3\pi}\left(\frac{3}{\lambda}\right)^{5/3}\frac{\Gamma(\frac{1}{3})[\Gamma(\frac{1}{6})]^2}{\tau^2}\left(1+\frac{1}{\sqrt{3}}\cos\left(\frac{2}{3}\lambda\tau^3-\frac{5\pi}{6}\right)\right).
\end{equation*}
For $n=1$, since ${}_2F_3(1,1;\frac{8}{6},\frac{9}{6},\frac{10}{6};-z)$ is a linear combination of $z^{-1}$ and $z^{-1}\ln z$ near $z=\infty$, $R_1(\tau)$ is a linear combination of $\tau^{-2}$ and $\tau^{-2}\ln\tau$. Hence, we have $R_2(\tau)\gg R_1(\tau)\gg  R_0(\tau)$ near $\tau=\infty$.

\end{appendix}

\end{document}